\documentclass[11pt]{article}

\usepackage[utf8]{inputenc}
\usepackage[T1]{fontenc}
\usepackage[english]{babel}
\usepackage{amsthm,amsmath,amssymb,verbatim}
\usepackage{color}
\usepackage{framed}
\usepackage{graphicx}
\usepackage{csquotes}
\usepackage{setspace}
\usepackage{wrapfig}
\usepackage{caption}
\usepackage{cite}

\newtheorem{theorem}{Theorem}[section]

\newtheorem{proposition}{Proposition}[section]
\newtheorem{definition}{Definition}[section]
\newtheorem{remark}{Remark}[section]
\newtheorem{example}{Example}[section]

\DeclareMathOperator{\mat}{Mat}
\DeclareMathOperator{\gal}{Gal}
\DeclareMathOperator{\rank}{rk}
\DeclareMathOperator{\nrm}{Nm}
\DeclareMathOperator{\Nrm}{N}
\DeclareMathOperator{\rnm}{nm}
\DeclareMathOperator{\tra}{Tr}
\DeclareMathOperator{\rtr}{tr}
\DeclareMathOperator{\diag}{diag}
\DeclareMathOperator{\id}{id}

\DeclareMathOperator*{\argmin}{arg\,min}

\DeclareMathOperator{\snr}{SNR}
\DeclareMathOperator{\vol}{vol}
\DeclareMathOperator{\Hom}{Hom}

\DeclareMathOperator{\ind}{ind}

\DeclareMathOperator{\Aut}{Aut}

\newcommand{\Q}{\mathbb{Q}}

\newcommand{\Z}{\mathbb{Z}}
\newcommand{\R}{\mathbb{R}}
\newcommand{\C}{\mathbb{C}}
\newcommand{\F}{\mathbb{F}}

\newcommand{\mb}[1]{\mathbf{#1}}
\newcommand{\mc}[1]{\mathcal{#1}}
\newcommand{\mf}[1]{\mathfrak{#1}}
\newcommand{\vl}[1]{\vol\left(#1\right)}
\newcommand{\rk}[1]{\rank\left(#1\right)}
\newcommand{\nm}[2]{\nrm_{#1}\left(#2\right)}

\newcommand{\tr}[2]{\tra_{#1}\left(#2\right)}

\title{On Fast-Decodable Algebraic Space--Time Codes}

\author{
Amaro Barreal \\
Aalto University \\
amaro.barreal@aalto.fi
\and
Camilla Hollanti \\
Aalto University \\
camilla.hollanti@aalto.fi
}

\date{}

\begin{document}

\maketitle
\newpage

\tableofcontents
\newpage

\section*{Abstract}

In the near future, the $5^{th}$ generation (5G) wireless systems will be established. They will consist of an integration of different techniques, including distributed antenna systems and massive multiple-input multiple-output systems, and the overall performance will highly depend on the channel coding techniques employed. Due to the nature of future wireless networks, space--time codes are no longer merely an object of choice, but will often appear naturally in the communications setting. However, as the involved communication devices often exhibit a modest computational power, the complexity of the codes to be utilised should be reasonably low for possible practical implementation. 

Fast-decodable codes enjoy reduced complexity of maximum-likelihood (ML) decoding due to a smart inner structure allowing for parallelisation in the ML search. The complexity reductions considered in this chapter are entirely owing to the algebraic structure of the considered codes, and could be further improved by employing non-ML decoding methods, however yielding suboptimal performance.

The aim of this  chapter is twofold. First, we provide a tutorial introduction to space--time coding and study powerful algebraic tools for their design and construction. Secondly, we revisit algebraic techniques used for reducing the worst-case decoding complexity of both single-user and multiuser space-time codes, alongside with general code families and illustrative examples.
\newpage

\section{Introduction}
\label{sec:intro}

Let us start this chapter by introducing, very briefly, the reader to the field of algebraic space--time coding. While there are various design criteria to be considered as well as a plethora of code constructions for a variety of different channel models and communications settings, we will here only review the developments most relevant to the rest of this chapter.  

The first space--time code, the Alamouti code \cite{alamouti:stc}, was introduced in 1998 and gave rise to a massive amount of research in the attempt to construct well-performing codes for various multi-antenna wireless communications settings. It was discovered that the code matrices constituting this particular code actually depict an algebraic structure known as the Hamiltonian quaternions, and by restriction to Lipschitz (\emph{i.e.}, integral) quaternions, the (unconstrained) code becomes a lattice. As Hamiltonian quaternions are the most popular example of a division algebra, this finding prompted the study of general division algebra space--time lattice codes \cite{sethuraman:stc, belfiore:quat}. 

The relevance of being division is related to achieving full diversity by maximising the rank of the code matrices \cite{tarokh:stc}. Soon it was noticed that by choosing the related field extensions carefully, one can achieve non-vanishing determinants (NVD) \cite{belfiore:quat} for the codewords, implying a non-vanishing coding gain \cite{tarokh:stc}. As the coding gain is inversely proportional to the decoding error probability, this in turn prevents the error probability from blowing up. A related notion, the diversity--multiplexing gain \cite{zheng:dmt} captures the tradeoff between the decay speed of the decoding error probability and available degrees of freedom. It is known that for symmetric systems, that is, with an equal number of transmit and receive antennas, full-rate space-time codes with the NVD property achieve the optimal tradeoff of the channel.

Several explicit constructions of space--time codes based on cyclic division algebras exist in the literature. For instance, the Perfect space-time codes and their generalisations \cite{belfiore:golden,oggier:perfect,kumar:genperf} provide orthogonal lattices for any number of antennas, whereas the maximal order codes \cite{hollanti:order1,hollanti:order2,vehkalahti:dense_mimo} optimise the coding gain, while giving up on the orthogonality of the underlying lattice.  

In the multiuser settings considered in this chapter, multiple users are communicating to a joint destination, with or without cooperating with each other. When cooperation is allowed, it is possible to take advantage of intermediate distributed relays which aid the active transmitter in the communication process. Various protocols exist for enabling this type of diversity --- the one considered here is the non-orthogonal half-duplex amplify-and-forward protocol, see \cite{yang:af}. The non-cooperative case is referred to as the multiple access channel (MAC), where users transmit signals independently of each other. Some algebraic MAC codes are presented in \cite{francis:2mac,francis:dmt_mac}, among others.  

One of the biggest obstacles in utilising space-time lattice codes and realising the theoretical promise of performance gains is their decoding complexity. Namely, maximum-likelihood (ML) decoding boils down to closest lattice point search, the complexity of which grows exponentially in the lattice dimension. More efficient methods exist, most prominently sphere decoding \cite{viterbo:sphere_decod}, which limits the search to a hypersphere of a given radius. However, the complexity remains prohibitive for higher dimensional lattices. To this end, several attempts have been made to reduce the ML decoding complexity. In principle, there are two ways to do this: either one can resort to reduced-complexity decoders yielding suboptimal performance, or try to build the code lattice in such a way that its structure naturally allows for parallelisation of the decoding process, hence yielding reduction in the dimensionality of the search. In this chapter, we are interested in the latter: we will show how to design codes that inherently yield reduced complexity thanks to a carefully chosen underlying algebraic structure. 

On our way to this goal, we will introduce the reader to the basics of lattices and algebraic number theory, to the extent that is relevant to this chapter. We will also lay out the typical channel models for the considered communications settings. Whenever we cannot explain everything in full detail in the interest of space, suitable references will be given for completeness. We assume the reader is familiar with basic abstract algebra and possesses some mathematical maturity, while assuming no extensive knowledge on wireless communications. 

The rest of the chapter is organised as follows. We begin in Section~\ref{sec:algebra} by familiarising the reader with the important notion of lattices and recall related results. Following a section introducing concepts and results from algebraic number theory, we study a particular class of central simple algebras, specifically cyclic division algebras, and their orders.
We then move on to providing a background in wireless communications in Section~\ref{sec:physical_layer}, introducing the well-known multiple-input multiple-output fading channel model and related performance parameters. As a coding technique employed in this multiple-antenna communications setup, we then introduce the main object of this chapter, space--time codes. We recall code design criteria, and furthermore show how codes can be constructed from cyclic division algebras. In Section~\ref{sec:reduced_ml}, maximum-likelihood decoding is introduced, and we discuss a possible decoding complexity reduction by algebraic means, defining the concept of fast-decodable space--time codes. The definition of fast decodability is then further refined, which allows us to consider more specific families of space--time codes with reduced decoding complexity. We further recall a useful iterative method for code construction. Finally, in Section~\ref{sec:codes} we consider two specific communication scenarios for which explicit construction methods which give rise to fast-decodable space--time codes can be recalled. 
\newpage

\section{Algebraic Tools for Space--Time Coding}
\label{sec:algebra}

While a tool employed for data transmission and thus falling into the area of communications theory, space--time codes are of a very mathematical nature. Indeed, design criteria derived for minimising the probability of incorrect decoding, which we will revisit in Section~\ref{subsubsec:stc_design}, can be met by ensuring certain algebraic properties of the underlying structure used for code construction. For this reason, we first devote a chapter to the mathematical notions needed for space--time code analysis and design.

We start with basic concepts and results about lattices, objects which are of particular interest as almost all space--time codes with good performance arise from lattice structures. This is both to ensure a linear structure -- a lattice is simply a free $\Z$-module, thus an abelian group -- as well as to avoid accumulation points at the receiver, to which end the discreteness property of a lattice is useful. Our main references for all lattice related results are \cite{conway:lattices, ebeling:lattices}.

In a successive section we then introduce relevant tools and objects from algebraic number theory, such as number fields, their rings of integers, and prime ideal factorisation. These tools will play a crucial role in the construction of space--time codes. As references serve \cite{milne:ant, neukirch:ant}.  

Most importantly, we finally introduce central simple algebras and their orders, the main objects that will determine the performance of the constructed codes. Over number fields, every central simple algebra is cyclic, and we study these in detail. We refer to \cite{milne:cft, oggier:cda, berhuy:book} for good general references.

\subsection{Lattices}
\label{subsec:lattices} 

We begin with the simplest definition of a lattice in the ambient space $\R^n$. 
\begin{definition}
	A \emph{lattice} $\Lambda \subset \R^n$ is the $\Z$-span of a set of vectors of $\R^n$, linearly independent over $\R$. 	
\end{definition}

Note that we do not require that the number of vectors spanning $\Lambda$ equals the dimension $n$. Indeed, any lattice is isomorphic to $\Z^t$ as \emph{groups} for $t \le n$. A lattice is thus a free abelian group of rank $\rk{\Lambda} = t$, and is called \emph{full-rank} or shortly \emph{full}, if the dimension and rank coincide, \emph{i.e.}, $t = n$. We give an alternative and useful group theoretic definition.
\begin{definition}
	A \emph{lattice} $\Lambda \subset \R^n$ is a discrete\footnote{By discrete we mean that the metric on $\R^n$ defines the discrete topology on $\Lambda$.} subgroup of $\R^n$. 
\end{definition}

A lattice $\Lambda \subseteq \R^n$ can hence be expressed as a set 
\begin{align*}
	\Lambda = \left\{\left.\mb{x} = \sum\limits_{i=1}^{t}{\mb{b}_i z_i} \right| z_i \in \Z\right\},
\end{align*}
with $\mb{b}_i \in \R^n$. We say that $\left\{\mb{b}_1,\ldots,\mb{b}_t\right\}$ forms a $\Z$-basis of $\Lambda$.

	We can conveniently define a \emph{generator matrix} and the corresponding \emph{Gram matrix} for $\Lambda$
\begin{align*}
	M_{\Lambda} = \begin{bmatrix} \mb{b}_1 & \cdots & \mb{b}_n \end{bmatrix}; \quad G_{\Lambda} = M_{\Lambda}^t M_{\Lambda},
\end{align*}
so that every element of $\Lambda$ can be expressed as $\mb{x} = M_{\Lambda}\mb{z}$ for some $\mb{z} \in \Z^n$. 

\begin{example}
	The simplest lattice is the integer lattice $\Z^n$ in arbitrary dimension $n \ge 1$. A generator and Gram matrix for $\Z^n$ is simply the $n\times n$ identity matrix. 
	
A more interesting example in dimension $n = 2$ is the \emph{hexagonal lattice} $A_2$. A $\Z$-basis for this lattice can be taken to be $\mb{b}_1 = (1,0)^t$ and $\mb{b}_2 = (-1/2,\sqrt{3}/2)^t$. A graphical representation of the lattice, as well as a generator and Gram matrix with respect to this basis are presented below in Figure 1. 
\vspace{.5cm}

\begin{minipage}[b]{0.58\textwidth}
	\includegraphics[width=.9\textwidth, height=.9\textwidth]{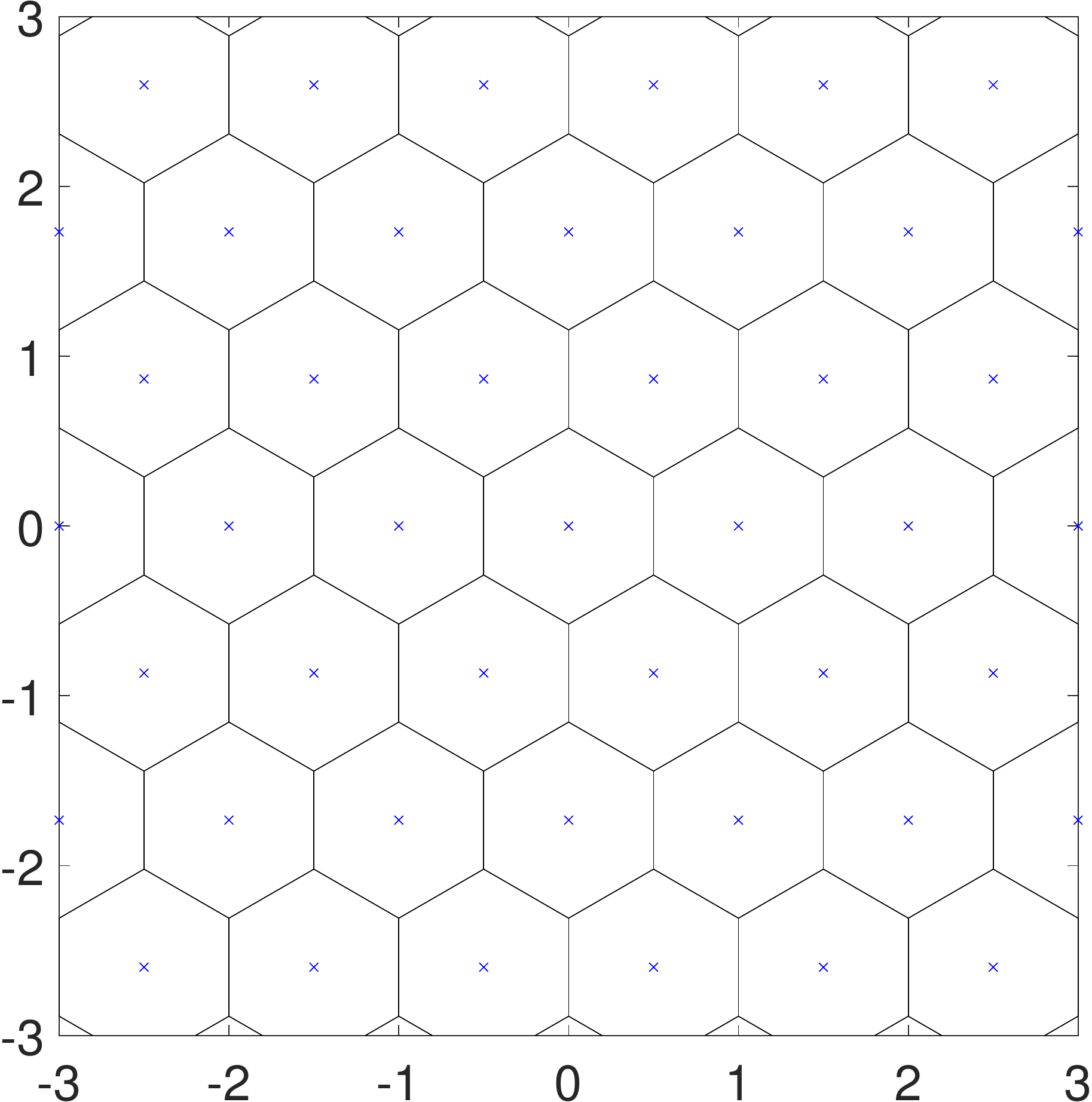}
	\captionof{figure}{The hexagonal lattice $A_2$.}
\end{minipage}
\begin{minipage}[t]{0.38\textwidth}
\vspace{-6.5cm}
	\begin{align*}
		M_{A_2} &= \begin{bmatrix} 1 & -\frac{1}{2} \\ 0 & \frac{\sqrt{3}}{2} \end{bmatrix}; \\[.5cm]
		G_{A_2} &= \begin{bmatrix} 1 & -\frac{1}{2} \\ -\frac{1}{2} & 1 \end{bmatrix}.
	\end{align*}
\end{minipage}
\end{example}
 
To each lattice $\Lambda$ we can associate its \emph{fundamental parallelotope}, defined as $\mc{P}_{\Lambda} := \left\{\left. M_{\Lambda}\mb{y}\right| \mb{y} \in [0,1)^n \right\}$. Note that we can recover $\R^n$ as a disjoint union of the sets $\mb{x} + \mc{P}_{\Lambda}$ for all $\mb{x} \in \Lambda$. 
Since $M_{\Lambda}$ contains a $\Z$-basis of $\Lambda$, any change of basis is obtained via an integer matrix with determinant $\pm 1$. Hence, the Lebesgue measure of $\mc{P}_{\Lambda}$ is invariant under change of basis. Thus, we define the \emph{volume} of a lattice $\Lambda \subset \R^n$ as the Lebesgue measure of its fundamental parallelotope, 
\begin{align*}
	\vl{\Lambda} := \vl{\mc{P}_{\Lambda}} = \sqrt{\det(G_{\Lambda})}.
\end{align*} 

We have defined a lattice to be a discrete subgroup of $\R^n$ and they are, by definition, free $\Z$-modules. It is however possible and often desirable to extend the definition to other rings and ambient spaces, such as the ring of integers of a number field, or an order in a cyclic division algebra. In this more general context, we define a lattice $\Lambda$ to be a discrete and finitely generated abelian subgroup of a real or complex ambient space $V$. In the previous derivations, we have set $V = \R^n$. Of interest for purposes of space--time coding is to consider lattices in $V = \mat(n,\C)$. In this case, we can also identify a full lattice in $V$ with a full lattice in $\R^{2n^2}$ via the $\R$-linear isometry 
\begin{align}
\label{eqn:isometry}
\begin{split}
	\iota: 	\mat(n,\C) &\to \R^{2n^2}, \\
	[\mb{u}_1,\ldots,\mb{u}_n] &\mapsto \left(\Re(u_{11}),\Im(u_{11}),\ldots,\Im(u_{1n}),\ldots,\Re(u_{nn}),\Im(u_{nn})\right)^t.
\end{split}
\end{align} 

We have $\|U\|_F = \|\iota(U)\|$, where $\|\cdot\|$ (resp. $\|\cdot\|_F$) denotes the Euclidean (resp. Frobenius) norm, and $\iota$ maps full lattices in $V$ to full lattices in the target Euclidean space. This map will be crucial for decoding considerations in later sections.

Let $\Lambda \subset \mat(n,\C)$ be a full lattice with $\Z$-basis $\left\{B_1,\ldots,B_n\right\}$, $B_i \in \mat(n,\C)$. A generator matrix and the corresponding Gram matrix for $\Lambda$ can be given as 
\begin{align*}
	M_{\Lambda} = \begin{bmatrix} \iota(B_1) & \cdots & \iota(B_n) \end{bmatrix}; \quad G_{\Lambda} = M_{\Lambda}^{t} M_{\Lambda} = \left(\Re(\tra(B_i^\dagger B_j))\right)_{i,j}.
\end{align*}
The volume of $\Lambda$ is the volume of the corresponding lattice $\iota(\Lambda)$ in $\R^{2n^2}$, \emph{i.e.}, $\vl{\Lambda} = \sqrt{\det(G_{\Lambda})}$. 

\begin{example}
	We exemplify the notion of a lattice in $\mat(n,\C)$ and corresponding vectorisation on the famous Alamouti code \cite{alamouti:stc}. As we shall see later, the Alamouti code is constructed from a lattice in $\mat(2,\C)$ corresponding to Hamiltonian (or more precisely Lipschitz) quaternions. More concretely, it is a finite subset 
	\begin{align*}
		\mc{X}_A \subset \left\{\left.\begin{bmatrix} x_1+ix_2 & -(x_3-ix_4) \\ x_3+ix_4 & x_1-ix_2 \end{bmatrix} \right| (x_1,\ldots,x_4) \in \Z^4 \right\}.
	\end{align*}
	
	A basis of the underlying lattice $\Lambda_A$ consists of the matrices 
	\begin{align*}
		B_1 = \begin{bmatrix} 1 & 0 \\ 0 & 1 \end{bmatrix}; \quad 
		B_2 = \begin{bmatrix} i & 0 \\ 0 & -i \end{bmatrix}; \quad 
		B_3 = \begin{bmatrix} 0 & -1 \\ 1 & 0 \end{bmatrix}; \quad 
		B_4 = \begin{bmatrix} 0 & i \\ i & 0\end{bmatrix}.
	\end{align*}
	
	Using the defined isometry $\iota$, we can identify $\Lambda_A$ with a lattice in $\R^8$, which we again denote by $\Lambda_A$, with generator and Gram matrix 
	\begin{align*}
		M_{\Lambda_A} = \begin{bmatrix} 
							1 & 0 & 0 & 0 \\
							0 & 1 & 0 & 0 \\
							0 & 0 & 1 & 0 \\
							0 & 0 & 0 & 1 \\
							0 & 0 & -1 & 0 \\
							0 & 0 & 0 & 1 \\
							1 & 0 & 0 & 0 \\
							0 & -1 & 0 & 0
						\end{bmatrix}; \qquad 
		G_{\Lambda_A} = \begin{bmatrix}
							2 & 0 & 0 & 0 \\ 
							0 & 2 & 0 & 0 \\
							0 & 0 & 2 & 0 \\
							0 & 0 & 0 & 2
						\end{bmatrix}
	\end{align*}
	
	The volume of this lattice is $\vl{\Lambda_A} = \sqrt{\det(G_{\Lambda_A})} = 4$.
\end{example}

\subsection{Algebraic Number Theory}
\label{subsec:ant}

In this section, we recall fundamental notions from algebraic number theory which are indispensable for space--time code constructions. We assume that the reader is familiar with basic Galois theory. 

Let $L/K$ be an arbitrary field extension. If we view $L$ as a vector space over $K$, we can define the degree of the field extension as the vector space dimension, that is, $\left[L:K\right] := \dim_{K}(L)$. An element $\alpha \in L$ is called \emph{algebraic} over $K$ if there exists a non-zero polynomial $f(x) \in K\left[x\right]$ such that $f(\alpha) = 0$, and the field extension $L/K$ is called \emph{algebraic} if all elements of $L$ are algebraic over $K$. Consider the homomorphism $\phi: K[x] \to L$, $f(x) \mapsto f(\alpha)$. Since $\alpha$ is algebraic, $\ker(\phi) \neq 0$, and can be generated by a single polynomial $m_{\alpha}(x)$, chosen to be monic of smallest degree admitting $\alpha$ as a root. We call this unique polynomial the \emph{minimal polynomial} of $\alpha$ over $K$.   

\begin{definition}
	An \emph{algebraic number field} is a finite extension of $\Q$. 
\end{definition}

\begin{example}
	The simplest example of a non-trivial number field is the Gaussian field $\Q(i)$, where $i=\sqrt{-1}$ is the imaginary unit. The minimal polynomial of $i \in \C$ over $\Q$ is given by $m_{i}(x) = x^2 + 1$.
\end{example}

We will henceforth consider $L/K$ to be an extension of algebraic number fields. In the above example, we constructed the field $\Q(i)$ by \emph{adjoining} an algebraic element $i \in \C$ to $\Q$. This is a more general phenomenon. 

\begin{theorem}[Primitive Element Theorem]
\label{thm:primitive}
	Let $L/K$ be a extension of number fields. Then, there exists an element $\alpha \in L$ such that $L = K(\alpha)$. 
\end{theorem}

We see that we can construct the field $L$ by adjoining the algebraic element $\alpha \in L$ to $K$ and, since $m_{\alpha}(x)$ is irreducible, we have the isomorphism
\begin{align*}
	L \cong K[x]/\langle m_{\alpha}(x)\rangle.
\end{align*}
It now becomes apparent that the degree of the field extension equals the degree of the minimal polynomial of the adjoined element, $\left[L:K\right] = \deg(m_{\alpha}(x))$. 

\begin{example}
	Consider the number field $K = \Q(\sqrt{2},\sqrt{3})$. We claim that $K = \Q(\sqrt{2}+\sqrt{3})$ and is hence generated by a single element. The inclusion $\Q(\sqrt{2}+\sqrt{3}) \subseteq K$ is trivial, as $\sqrt{2}+\sqrt{3} \in \Q(\sqrt{2},\sqrt{3})$. For the reverse inclusion, it suffices to express $\sqrt{2}$ and $\sqrt{3}$ in terms of elements of $\Q(\sqrt{2}+\sqrt{3})$. Note that as $(\sqrt{2}+\sqrt{3})^2 = 5+2\sqrt{6}$ it follows that $\sqrt{6} \in \Q(\sqrt{2}+\sqrt{3})$, and we have 
	\begin{align*}
		\sqrt{2} = \frac{2+\sqrt{6}}{\sqrt{2}+\sqrt{3}}; \quad \sqrt{3} = \frac{3+\sqrt{6}}{\sqrt{2}+\sqrt{3}}.
	\end{align*}
	
	This shows that $\Q(\sqrt{2},\sqrt{3}) = \Q(\sqrt{2}+\sqrt{3})$. The minimal polynomial of $\alpha := \sqrt{2}+\sqrt{3}$ is $m_{\alpha}(x) = x^4-10x^2+1$, and we see that $\Q(\alpha)$ is an extension of degree $4$. 
\end{example}

We now define a very important ring associated with a number field $K$.
\begin{definition}
	Let $K$ be a number field. The integral closure of $\Z$ in $K$ is a ring, called the \emph{ring of integers} $\mc{O}_K$ of $K$. It consists of all the elements $\alpha \in K$ for which $m_{\alpha}(x)\in\Z[x]$.
	We call any element $\alpha \in \mc{O}_K$ an \emph{algebraic integer}.  
\end{definition}

\begin{example}
	Consider the field extension $\Q(i)/\Q$. The ring of integers of $\Q(i)$ is precisely $\Z[i]$. It is however not always true that $\mc{O}_{K(\alpha)} = \Z[\alpha]$. Consider for example $\Q(\sqrt{5})/\Q$. We have that $\Z[\sqrt{5}]$ is composed of algebraic integers, but $\Z[\sqrt{5}] \neq \mc{O}_K$. For example, the element $\frac{1+\sqrt{5}}{2}$ is a root of the polynomial $x^2-x-1$, but $\frac{1+\sqrt{5}}{2} \notin \Z[\sqrt{5}]$. In fact, it turns out that $\mc{O}_K = \Z\left[\frac{1+\sqrt{5}}{2}\right]$. 
\end{example}

It is clear that $\alpha \in K$ is an algebraic integer if and only if $m_{\alpha}(x) \in \Z[x]$. Further, the field of fractions of $\mc{O}_K$ is precisely $K$. 
In the above examples, the ring of integers $\mc{O}_K = \Z[\theta]$ admits a $\Z$-basis $\left\{1,\theta\right\}$. In fact, we have the following result.
\begin{theorem}
	Let $K$ be a number field of degree $n$. The ring of integers $\mc{O}_K$ of $K$ is a free $\Z$-module of rank $n$.
\end{theorem} 

As a consequence, the ring of integers $\mc{O}_K$ admits an integral basis over $\Z$, that is, a basis as a $\Z$-module. Given an extension $L/K$ of number fields, it is however not true in general that the ring of integers $\mc{O}_L$ is a free $\mc{O}_K$-module. This holds for instance if $\mc{O}_K$ is a principal ideal domain (PID). We will be considering extensions of $\Q$ and $\Q(i)$, hence circumventing this problem\footnote{The practical reason behind this choice is that the popular modulation alphabets, referred to as pulse amplitude modulation (PAM) and quadrature amplitude modulation (QAM), correspond to the rings of integers of these fields.}.  

Consider a number field $K$ of degree $n$ over $\Q$. We fix compatible embeddings of $K$ into $\C$, and identify the field with its image under these embeddings. More precisely, there exist exactly $n$ pairwise distinct embeddings $\sigma_i: K \to \C$, forming the set $\Hom_{\Q}(K,\C) = \left\{\sigma_1,\ldots,\sigma_n\right\}$. 

We split the embeddings into those whose image is real or complex, respectively. More concretely, let $\sigma_1,\ldots,\sigma_r: K \to \R$, and $\sigma_{r+1},\ldots,\sigma_n : K \to \C$. Note that the embeddings with complex image come in conjugate pairs, of which there are exactly $s := \frac{n-r}{2}$. We call the tuple $(r,s)$ the \emph{signature} of the number field $K$. 

We can use the embeddings to define two important functions, namely the norm and trace of elements in $K$. For each $\alpha \in K$, consider the induced homomorphism $\varphi_{\alpha}: K \to K$, where for all $\beta \in K$, we have $\varphi_{\alpha}(\beta) = \alpha \beta$. By fixing a basis of $K$ over $\Q$, $\varphi_{\alpha}$ can be represented by a matrix $A_{\alpha}$. This is referred to as the \emph{left regular representation}. 
\begin{definition}
 Let $K$ be a number field of degree $n$, and let $\alpha \in K$. The \emph{norm} and \emph{trace} of $\alpha$, respectively, are defined as
\begin{align*}
	\nm{K}{\alpha} = \det(A_{\alpha}) = \prod\limits_{i=1}^{n}{\sigma_i(\alpha)}; \qquad \tr{K}{\alpha} = \tra(A_{\alpha}) = \sum\limits_{i=1}^{n}{\sigma_i(\alpha)}.
\end{align*}
\end{definition}

We note that the norm and trace are generally rational elements. When $\alpha \in \mc{O}_K$, however, we have $\nm{K}{\alpha}, \tr{K}{\alpha} \in \Z$. 

\begin{definition}
	Let $K$ be a number field of degree $n$, with ring of integers $\mc{O}_K$, and let $\left\{b_1,\ldots,b_n\right\}$ be an integral basis of $\mc{O}_K$. The \emph{discriminant} of $K$ is the well-defined integer 
	\begin{align*}
		d_K &= \det\left(\begin{bmatrix} \tr{K}{b_1 b_1} & \cdots & \tr{K}{b_1 b_n} \\ \vdots & \ddots & \vdots \\ \tr{K}{b_n b_1} & \cdots & \tr{K}{b_n b_n}\end{bmatrix}\right) \\
		    &= \det\left(\begin{bmatrix} \sigma_1(b_1) & \cdots & \sigma_1(b_n) \\ \vdots & \ddots & \vdots \\ \sigma_n(b_1) & \cdots & \sigma_n(b_n)\end{bmatrix}\right)^2.
	\end{align*}
\end{definition}
The discriminant $d_K$ is independent of the choice of basis, and hence an invariant of the number field.

\begin{example}
	Consider the number field $K = \Q(\sqrt{-5})$, with ring of integers $\mc{O}_K = \Z[\sqrt{-5}]$. As $K$ is a degree-$2$ extension of $\Q$, and generated by a complex element, we have that its signature is $(r,s) = (0,1)$. A representative of the pair of complex embeddings is given by $\sigma_1: \sqrt{-5} \mapsto -\sqrt{-5}$, and the complex conjugate $\sigma_2$ is simply the identity. 
	
	Given an element $\alpha = x_0+\sqrt{-5}x_1 \in K$, the norm and trace of $\alpha$ can be computed to be 
	\begin{align*}
		\nm{K}{\alpha} = \sigma_1(\alpha)\sigma_2(\alpha) = x_0^2+5x_1^2; \quad \tr{K}{\alpha} = \sigma_1(\alpha)+\sigma_2(\alpha) = 2x_0.
	\end{align*}
	
	Moreover, we can compute the discriminant of $K$ by choosing a basis $\left\{1,\sqrt{-5}\right\}$ of $\mc{O}_K$ and computing the determinant 
	\begin{align*}
		d_K = \det\left(\begin{bmatrix} 1 & -\sqrt{-5} \\ 1 & \sqrt{-5} \end{bmatrix} \right)^2 = -20.
	\end{align*}
\end{example}

The motivation for studying number fields has its origins in the factorisation of integers into primes. In the ring $\Z$, prime and irreducible elements coincide, and every natural number factors uniquely into prime numbers. By generalising the ring $\Z$ to the ring of integers $\mc{O}_K$ of a number field, unique factorisation into prime elements is no longer guaranteed. However, the underlying structure of the ring $\mc{O}_K$ allows for a generalisation of unique factorisation by making use of ideals, instead of elements. 

Let $K$ be a number field of degree $n$, and $\mf{a} \subset \mc{O}_K$ a non-zero ideal. Then $\mf{a}$ factors  into a product of prime ideals, unique up to permutation,
\begin{align*}
	\mf{a} = \prod\limits_{i=1}^{g}{\mf{p}_i^{e_i}},
\end{align*} 
where $e_i > 0$. We define the \emph{norm} of the ideal $\mf{a}$ as the cardinality of the finite ring $\Nrm(\mf{a}) := \left|\mc{O}_K/\mf{a}\right|$. The ideal norm extends multiplicatively, and moreover $\Nrm(\mf{a}) \in \mf{a}$. Consequently, if $\Nrm(\mf{a})$ is prime, then $\mf{a}$ is a prime ideal. More importantly, if $\Nrm(\mf{a}) = p_1^{e_1}\cdots p_k^{e_k}$ is the prime factorisation, then it is clear that as $\mf{a}$ divides $\Nrm(\mf{a})\mc{O}_K$, every prime divisor of $\mf{a}$ is a prime divisor of $p_i \mc{O}_K$ for some $i$.  

\begin{remark}
	If all prime divisors of $p\mc{O}_K$ are known for all primes $p \in \Z$, then all ideals of $\mc{O}_K$ are known. 
\end{remark}

Let $\mf{p} \subset \mc{O}_K$ be a prime ideal. Then $\mf{p}\cap \Z = p\Z$ is a prime ideal of $\Z$, $p$ prime. We can hence write 
\begin{align*}
	p\Z = \mf{p}^{e}\mf{p}_2^{e_2} \cdots \mf{p}_k^{e_k}
\end{align*} 
for $\mf{p}_i$ distinct prime ideals of $\mc{O}_K$. The number $e = e(\mf{p}/p\Z)$ is referred to as the \emph{ramification index} of $p\Z$ at $\mf{p}$. We further define the \emph{residue class degree} as the integer $f \ge 1$ which satisfies $\Nrm(\mf{p}) = p^f$.

\begin{example}
	Consider $K = \Q(i)$, and let $p>2 $ be a rational prime. We want to study the factorisation of $p$ in $\mc{O}_K = \Z[i]$. By norm considerations, as $\Nrm(p\Z[i]) = p^2$, we have that $p$ can either remain prime in $\Z[i]$, or be the product of two prime ideals. On the other hand, we know that $p\Z[i]$ is prime if and only if $\Z[i]/\langle p \rangle$ is a field. In fact, 
	\begin{align*}
		\Z[i]/\langle p \rangle \cong \Z[x]/\langle p, x^2+1\rangle \cong \F_p[x]/\langle x^2+1\rangle,
	\end{align*}
	so that the residue class degree is $f = 2$. This quotient is a field precisely when $x^2+1$ is irreducible. This is the case for $p \not\equiv 1 \bmod{4}$.
	
For the case $p \equiv 1 \bmod{4}$, we can factor $x^2+1 = (x-a)(x-b)$, and we get a factorisation $p\Z[i] = (i-a)(i-b)$. 
\end{example}

\subsection{Central Simple Algebras}
\label{subsec:csa}

Let $K$ be a field, and $\mc{A}$ a finite-dimensional associative $K$-algebra, \emph{i.e.}, a finite-dimensional $K$-vector space together with a $K$-bilinear product. The algebra is \emph{simple}, if it contains no non-trivial two-sided ideals, and moreover \emph{central} if its centre is precisely $K$. The algebra is a \emph{division algebra} if all of its non-zero elements are invertible. We have the following important theorem, which is a simplified version of a more general result. 

\begin{theorem}[Wedderburn]
	Every simple $K$-algebra is isomorphic to $\mat(n,D)$ for some uniquely determined $n$ and some division $K$-algebra $D$, unique up to isomorphism.
\end{theorem}

If $\mc{A}$ is a simple $K$-algebra and $D$ is the division algebra from the above theorem, we denote by $\ind(\mc{A}) = \sqrt{\left[D:K\right]}$ the \emph{index}, and by $\deg(\mc{A}) = \sqrt{\left[\mc{A}:K\right]}$ the \emph{degree} of the algebra. $\mc{A}$ is a division algebra if and only if $\ind(\mc{A}) = \deg(\mc{A})$.

If $K$ is a number field, every $K$-central simple algebra is \emph{cyclic}, and vice versa. This family of central simple algebras has been widely used for space--time coding since the work \cite{sethuraman:stc}. We start with the special case of cyclic algebras of degree $2$, also known as \emph{quaternion algebras}.
\begin{definition}
	Let $K$ be a field, and $a,\gamma \in K^\times$ not necessarily distinct. A \emph{quaternion algebra} is a set of all expressions 
	\begin{align*}
		(a,\gamma)_K := \left\{\left. x = x_0+ix_1+jx_2+kx_3 \right| x_i \in K \right\},
	\end{align*}
	where the basis elements  satisfy the rules
	\begin{align*}
		i^2 = a, \quad j^2 = \gamma, \quad ij = -ji = k.
	\end{align*}
\end{definition} 

\begin{example}
	The most famous example is the set of \emph{Hamiltonian quaternions}, which can be defined as $\mathbb{H} = (-1,-1)_\R$. An element $x \in \mathbb{H}$ is of the form $x = x_0+ix_1+jx_2+kx_3$ with $(x_0,x_1,x_2,x_3) \in \R^4$, $i^2 = j^2 = -1$ and $ij = -ji = k$. 
\end{example}

For quaternion algebras, we have the following deep and important classification result.

\begin{theorem}
	Let $(a,\gamma)_K$ be a quaternion algebra. We have two possibilities.
	\begin{itemize}
		\item[a)] $(a,\gamma)_K$ is a division algebra.
		
		\item[b)] $(a,\gamma)_K \cong \mat(2,K)$. 
	\end{itemize}
\end{theorem}

We can determine which of the cases apply by means of a simple quaternary quadratic form. To be more precise, consider an element $x = x_0 + ix_1 + jx_2 + kx_3 \in (a,\gamma)_K$, and define the \emph{norm} of $x$ as 
\begin{align*}
	\nrm(x) = xx^\ast = x_0^2-ax_1^2-\gamma x_2^2+a\gamma x_3^2,
\end{align*}
where $x^\ast = x_0 - ix_1 - jx_2 - kx_3$ is the conjugate of $x$ (note that this is also the complex conjugation when restricted to $\mathbb{C}$). Then, the quaternion algebra $(a,\gamma)_K$ is division if and only if $\nrm(x) = 0$ implies $x = 0$. 

\begin{example}
	Recall the set of Hamiltonian quaternions $\mathbb{H}$. The norm of an element $x = x_0+ix_1+jx_2+kx_3 \in \mathbb{H}$ is $\nrm(x) = x_0^2+x_1^2+x_2^2+x_3^2 \ge 0$. As $x_i \in \R$, we have equality if and only if $x = 0$. Hence, $\mathbb{H}$ is a division algebra.  
\end{example}

A quaternion algebra is a degree-$4$ vector space over the centre $K$. They are a special case of the more general cyclic algebras, a family of central simple algebras which we study in the sequel.

\begin{definition} 
	Let $L/K$ be a degree-$n$ cyclic Galois extension of number fields, and denote by $\langle \sigma \rangle = \gal\left(L/K\right)$ its Galois group. A \emph{cyclic algebra} is a tuple 
\begin{align*}
	\mc{C} = (L/K,\sigma,\gamma) := \bigoplus\limits_{i=0}^{n-1}{e^i L},
\end{align*}
where $e^n = \gamma \in K^\times$ and multiplication satisfies $le = e\sigma(l)$ for all $l \in L$.

The algebra $\mc{C}$ is called a \emph{cyclic division algebra} if it is division. 
\end{definition}

The usefulness of cyclic division algebras for purposes of space--time coding starts with the existence of a matrix representation of elements of the algebra. To be more precise, each element $x = \sum_{i=0}^{n-1}{e^i x_i} \in \mathcal{C}$ infers for all $y \in \mc{C}$ a right $L$-linear map $\rho: x \mapsto xy$, which is referred to as the \emph{left-regular representation} of the algebra, and describes left multiplication with $x$. We can define a matrix associated with $\rho$, given by 
\begin{align*}
	x \mapsto \rho(x) := \begin{bmatrix} 
				x_0 & \gamma\sigma(x_{n-1}) & \gamma\sigma^2(x_{n-2}) & \cdots & \gamma\sigma^{n-1}(x_1) \\ 
				x_1 & \sigma(x_0) & \gamma\sigma^2(x_{n-1}) & & \gamma\sigma^{n-1}(x_2) \\
				\vdots & & \vdots & & \vdots \\
				x_{n-2} & \sigma(x_{n-3}) & \sigma^2(x_{n-4}) & & \gamma\sigma^{n-1}(x_{n-1}) \\
				x_{n-1} & \sigma(x_{n-2}) & \sigma^2(x_{n-3}) & \cdots & \sigma^{n-1}(x_0) \end{bmatrix}. 
\end{align*}

\begin{example} 
 Let us consider again the Hamiltonian quaternions. Using the above notation, we write $e=j$ and  
$$
 \mathbb{H}=(\mathbb{C}/\mathbb{R},\sigma=*,\gamma=-1)=\mathbb{C}\oplus j\mathbb{C},
 $$
 with $cj=jc^*$ for all $c\in \mathbb{C}$ and $j^2=\gamma=-1$.
 Note that we have intentionally chosen to represent $\mathbb{H}$ as a right vector space in order to be compatible with the left regular representation. 
 
 Let now $x=x_0+jx_1$ with $x_0,x_1\in \mathbb{C}$. If we multiply the basis elements $\{1,j\}_{\mathbb{C}}$ from the left by $x$, we get
 \begin{align*}
 x\cdot 1 &= x_0+jx_1\,,\\
 x\cdot j &= (x_0+jx_1)j=x_0j+jx_1j=jx_0^*+j^2x_1^*=-x_1^*+jx_0^*\,.\\
 \end{align*}
  In a matrix form, we have
  $$
  x \mapsto \rho(x) = \begin{bmatrix}
  x_0 & -x_1^*\\
  x_1 & x_0^*\\
  \end{bmatrix}.
  $$
  Note that this matrix corresponds to the Alamouti code.
  \end{example}
  
\begin{example}
	Let $L/K$ be a number field extension of degree $n = 3$. Then, we can pick a basis $\left\{1,e,e^2\right\}$ of a cyclic algebra $\mc{C}$ over its maximal subfield $L$, where $e^3 = \gamma \in K^\times$. Let $x = x_0+ex_1+e^2 x_2$, and consider left multiplication.
	Similarly as above,
	\begin{align*}
	 x\cdot 1 &= x_0+ex_1+e^2 x_2\,,\\
	 x\cdot e &= (x_0+ex_1+e^2 x_2)e=e\sigma(x_0)+e^2\sigma(x_1)+e^3 \sigma(x_2)\\
	 &=	\gamma\sigma(x_2)+e\sigma(x_0)+e^2\sigma(x_1)\,,\\
	 x\cdot e^2 &= (x_0+ex_1+e^2 x_2)e^2=e^2\sigma(x_0)+e^3\sigma(x_1)+e^4\sigma(x_2)\\
	 &=\gamma\sigma(x_1)+\gamma e \sigma(x_2)+e^2\sigma(x_0)\,.\\
\end{align*}
	We see that in this basis, left multiplication by $x$ can be represented by the matrix 
	\begin{align*}
		\rho(x) = \begin{bmatrix} x_0 & \gamma\sigma(x_2) & \gamma\sigma^2(x_1) \\ x_1 & \sigma(x_0) & \gamma\sigma^2(x_2) \\ x_2 & \sigma(x_1) & \sigma^2(x_0) \end{bmatrix}
	\end{align*}
\end{example}

We close this section by recalling how to ensure that a cyclic algebra $(L/K,\sigma,\gamma)$ is division by means of the element $\gamma \in K^\times$. The result is a simple corollary to a result due to A. Albert. 

\begin{theorem}
\label{thm:non_norm}
	Let $\mc{C} = (L/K,\sigma,\gamma)$ be a cyclic algebra. If $\gamma^{n/p}$ is not a norm of some element of $L$ for all prime divisors $p$ of $n$, then $\mc{C}$ is division. 
\end{theorem}

\subsubsection{Orders}
\label{subsubsec:orders}

Given a number field $K$, the collection of integral elements form the ring of integers $\mc{O}_K$ of $K$. This ring is the unique \emph{maximal order} of $K$, a concept which we will now recall in a more general context. 

\begin{definition}
Let $\mc{C} = (L/K,\sigma,\gamma)$ be a cyclic division algebra. An $\mc{O}_K$-\emph{order} $\Gamma$ in $\mc{C}$ is a subring of $\mc{C}$ sharing the same identity as $\mc{C}$ and such that $\Gamma$ is a finitely generated $\mc{O}_K$-module which generates $\mc{C}$ as a linear space over $K$. 

An order is \emph{maximal} if it is not properly contained in another order of $\mc{C}$.
\end{definition}

Every order of a cyclic division algebra is contained in a maximal order. Within a number field $K$, the ring of integers $\mc{O}_K$ is integrally closed and the unique maximal order of $K$. In general, a maximal order $\Gamma$ of $\mc{C}$ is not integrally closed, and a division algebra $\mc{C}$ may contain multiple maximal orders. In contrast, the following special order is often of interest due to its simple structure. It is in fact the initial source for space--time codes with non-vanishing determinants.
\begin{definition}
	Let $\mc{C} = (L/K,\sigma,\gamma)$ be a cyclic division algebra. The \emph{natural order} of $\mc{C}$ is the $\mc{O}_K$-module
	\begin{align*}
		\Gamma_{\mathrm{nat}} := \bigoplus\limits_{i=0}^{n-1}{e^i \mc{O}_L}.
	\end{align*}
\end{definition}
Note that $\Gamma_{\mathrm{nat}}$ is not closed under multiplication unless $\gamma \in \mc{O}_K$. 

\begin{remark}\label{rednorm}
Given a cyclic division algebra $\mc{C} = (L/K,\sigma,\gamma)$ and an element $c \in \Gamma$, where $\Gamma \subset \mc{C}$ is an order, we define the \emph{reduced norm} $\rnm(c) = \det(\rho(c))$ and \emph{reduced trace} $\rtr(c) = \tra(\rho(c))$. These are elements of the ring of integers of the centre, \emph{i.e.}, $\rnm(c), \rtr(c) \in \mc{O}_K$. Consequently, for $K = \Q$ or $K$ imaginary quadratic, we have $\rnm(c) \ge 1$ for any non-zero $c \in \Gamma$, an observation which is crucial for achieving the \emph{non-vanishing determinant} property (cf. Section~\ref{subsubsec:stc_design}).
\end{remark}
\newpage

\section{Physical Layer Communications}
\label{sec:physical_layer}

In this section, we study the characteristics and properties of a wireless channel, discussing various methods for combating the effects of fading and noise.

\subsection{Rayleigh Fading MIMO Channel}
\label{subsec:mimo}

In a wireless environment, in contrast to wired channels, the signal traverses several  different paths between a transmitter and receiver. Consequently, different versions of the signal distorted by (independent) environmental effects will come together at the receiver, causing a superimposed channel output. Together with dissipation effects caused by urban scatterers as well as interference, the signal experiences \emph{fading}, and various statistical models exist to describe these phenomena. Here, we consider the widely used \emph{Rayleigh fading} channel model. In addition, thermal noise at the receiver further distorts the channel output. 

To be more precise, assume a single source, equipped with $n_t \ge 1$ transmit antennas, and a single destination, with $n_r \ge 1$ receive antennas. If $n_t, n_r \ge 2$ we refer to the setup as the \emph{multiple-input multiple-output} (MIMO) model, while the case $(n_t,n_r) = (1,1)$ is termed the \emph{single-input single-output} (SISO) channel model. The mixed cases $(n_t = 1, n_r > 1)$ and $(n_t > 1, n_r=1)$ are the SIMO and MISO channel setups, respectively.  

Consider a channel between $n_t$ transmit antennas and $n_r$ receive antennas. The wireless channel is modelled by a random matrix 
\begin{align*}
	H = \begin{bmatrix} 
		h_{11} & h_{12} & \cdots & h_{1n_t} \\ 
		h_{21} & h_{22} &  & h_{2n_t} \\
		\vdots & & \ddots & \vdots \\
		h_{n_r 1} & h_{n_r 2} & \cdots & h_{n_r n_t}
	\end{bmatrix} \in \mat(n_r \times n_t,\C), 
\end{align*}	
We assume that the channel remains static for $T\geq n_t$ time slots and then changes independently of its previous state, and refer to $T$ as the \emph{channel delay} or \emph{channel coherence time}. Each of the entries $h_{ij}$ of $H$ denotes the path gain from transmit antenna $j$ to receive antenna $i$. They are modelled as complex variables with i.i.d. normal distributed real and imaginary parts, 
\begin{align*}
	\Re(h_{ij}), \Im(h_{ij}) \sim \mc{N}(0,\sigma_h^2),
\end{align*} 
yielding a Rayleigh distributed envelope 
\begin{align*}	
	|h_{ij}| = \sqrt{\Re(h_{ij})^2 + \Im(h_{ij})^2} \sim \mathrm{Ray}(\sigma_h) 
\end{align*}
with scale parameter $\sigma_h$, which gives this fading model its name. 

The additive noise is modelled by a matrix $N \in \mat(n_r\times T,\C)$ with i.i.d. complex Gaussian entries with finite variance $\sigma_n^2$. To combat the destructive effects of fading, the transmitter encodes its data into a codeword matrix $X \in \mat(n_t\times T, \C)$. Each column $\mb{x}_i$ of $X$ corresponds to the signal vector transmitted in each time slot, across the available transmit antennas. If we denote each column of the noise matrix $N$ by $\mb{n}_i$, the received signal at each time slot $1 \le i \le T$ is given by the channel equation 
\begin{align*}
	\mb{y}_i = H\mb{x}_i + \mb{n}_i.
\end{align*}
We assume that the destination waits for the $T$ subsequent transmissions before starting any decoding process. As usual, we assume perfect channel state information at the receiver, while the transmitter only has statistical channel information. The channel is supposed to remain fixed during the entire transmission process, and hence we can summarise the overall channel equation in a compact form to read 
\begin{align*}
	Y = HX + N.
\end{align*}

Thus, by allowing the use of multiple antennas at the transmitter and/or receiver, we have created \emph{spatial diversity}. By ensuring a separation of the antennas by at least half the used wavelength, the multiple signals will fade independently of each other. On the other hand, the transmission over multiple time slots enables \emph{temporal diversity}, providing copies of the signal at the receiver. 

The physical conditions in an actual wireless channel are rapidly changing. Consequently, the comparison in performance of two different codes needs to be considered with respect to a standardised quantity. We define the \emph{signal-to-noise ratio} ($\snr$) at the receiver as the ratio of the received signal power to noise power, that is, 
\begin{align*}
	\snr = \frac{\mathbb{E}\left[\|HX\|^2\right]}{\mathbb{E}\left[\|N\|^2\right]}.
\end{align*}

\subsubsection{Performance Parameters of a Wireless Channel}
\label{subsubsec:channel_performance}

Consider a MIMO channel with $n_t$ transmit antennas and $n_r$ receive antennas. The first quantity that we need to mention is the \emph{capacity} of the channel.

\begin{definition}
	Assume that the receiver knows the realisation of the channel matrix $H$. For a fixed power constraint on the channel input, the \emph{capacity} of a MIMO channel is the upper bound on the mutual information between the channel input and output, given the channel realisation.
\end{definition}

As the capacity depends on the channel matrix, it needs to be viewed as a random variable. The ergodic capacity of a MIMO channel is given by 
\begin{align*}
	C_H = \mathbb{E}_H\left[\log\det\left(I_{n_r}+\frac{\snr}{n_t}H^\dagger H\right)\right].
\end{align*} 
Recently, the authors in \cite{luzzi:capacity} gave criteria for algebraic space--time codes from division algebras to achieve the channel capacity up to a constant gap. 

Equivalently we can interpret the capacity of the channel as the upper bound on the amount of information that can be transmitted, so that the probability of error can be maintained arbitrarily low. At high $\snr$, the capacity of the channel scales with the number of antennas. More specifically, an $\snr$ increase of 3dB results in an increase in capacity by $\min\left\{n_t,n_r\right\}$. 

We now define two quantities which allow us to compare different coding strategies for the MIMO channel. 
\begin{definition}
Consider a MIMO channel.  
	\begin{itemize}
		\item[i)] The \emph{diversity gain} of a coding strategy is the asymptotic slope of the corresponding error probability curve with respect to the $\snr$ in a $\log-\log$ scale. 
		\item[ii)] The \emph{coding gain} measures the difference in $\snr$ required for two different full-diversity coding strategies to achieve the same error probability.
	\end{itemize}
\end{definition}

\subsection{Space--Time Codes}
\label{subsec:stc}

This section introduces the main object of the survey: space--time codes. These codes are tailor-made for MIMO communications. We start with basic definitions and relate the enabled spatial and temporal diversity to the matrix structure of space--time codewords. 

In the first subsection, the basic code design criteria for minimising the probability of incorrect decoding are derived. While the design criteria are independent of the actual code construction method and hold for any matrix codebook, various results are then introduced exposing how the criteria can be met by purely algebraic means. Hence, it becomes clear which properties the underlying structures should exhibit in order to construct well-performing codes. 

After this, we utilise the algebraic tools introduced in Section~\ref{sec:algebra} in order to construct space--time codes meeting the derived criteria.

\subsubsection{Design Criteria for Space--Time Codes}
\label{subsubsec:stc_design}

Recall the Rayleigh fading $n_t\times n_r$ MIMO channel model with channel coherence time $T$. 
We have seen that the codewords $X$ need to be taken from some collection of matrices $\mc{X} \subset \mat(n_t\times T,\C)$. Very naively, and this is our first definition, we simply define a code to be a finite collection of such matrices. 
\begin{definition}
	Let $\mf{C} \subset \R^\times$ be a finite subset and $k\in\Z_+$. A \emph{space--time code} is the image of an injective map $\phi: \mf{C}^k \to \mat(n_t\times T,\C)$. 
\end{definition}

Having no structure, however, may lead to accumulation of the received signals. To circumvent this problem, forcing a discrete and linear structure on the code is helpful, \emph{e.g.}, a lattice structure. We give the more specialised definition of \emph{linear} space--time codes, which we will consider henceforth.
\begin{definition}
\label{def:stc}
	Let $\left\{ B_i \right\}_{i=1}^{k}$ be an independent set of matrices in $\mat(n_t\times T,\C)$. A \emph{linear space--time block code} of rank $k$ is a set of the form
\begin{align*}
	\mc{X} = \left\{\left. \sum\limits_{i=1}^{k}{B_i s_i} \right| s_i \in S \right\},
\end{align*}
where $S \subset \Z$ is a finite \emph{signalling alphabet}. 

If the matrices $\left\{ B_i \right\}_{i=1}^{k}$ form a basis of a \emph{lattice} $\Lambda \subset \mat(n_t\times T, \C)$, then $\mc{X}$ is called a \emph{space--time lattice code} of rank $k = \rk{\Lambda}\leq 2n_tT$. 
\end{definition}

We henceforth refer to such a code $\mc{X}$ simply as a space--time code. As the transmit power consumption is directly related to the Frobenius norm of the transmitted codeword, the finite codebook is usually carved out to consist of a desired number of lattice elements with smallest possible Frobenius norms\footnote{The smallest Frobenius norms correspond to the shortest Euclidean norms of the vectorised matrices. Directly, this would mean spherical constellation shaping. However, it is often more practical to consider a slightly more relaxed cubic shaping. This is the case in particular when the lattice in question is orthogonal, as then the so-called Gray-mapping \cite{gray:label} will give an optimal bit labelling of the lattice points.}.  

The \emph{code rate} of $\mc{X}$ is defined as $R = k/T$ real symbols per channel use\footnote{In the literature, the code rate is often defined in complex symbols per channel use. We prefer using real symbols, as not every code admits a Gaussian basis, while every lattice has a $\Z$-basis.}. For $n_r$ receive antennas, the code is said to be \emph{full rate} if $R = 2 n_r$. Here, full rate is defined as the maximum rate that still maintains the discrete structure at the receiver and allows for linear detection methods such as sphere-decoding \cite{viterbo:sphere_decod}. 

Consider a space--time code $\mc{X}$, and let $X \in \mc{X}$ be the transmitted codeword. A receiver observes its channel output $Y$ and, as it is assumed to know the channel $H$ and the noise is zero-mean, decodes a maximum-likelihood estimate of the transmitted codeword by computing 
\begin{align}
\label{eqn:ml_decod}
	\hat{X} = \argmin\limits_{X \in \mc{X}}{\|Y-HX\|_F^2}.
\end{align}
 
The probability $\Pr(X \to X')$ that a codeword $X' \neq X$ is decoded when $X$ was transmitted is asymptotically upper bounded with increasing $\snr$
  as 
\begin{align*}
	\Pr(X \to X') \le \left(\det\left((X-X')(X-X')^\dagger\right)\snr^{n_t}\right)^{-n_r}.
\end{align*}

From this upper bound, two design criteria can be derived \cite{tarokh:stc}. The \emph{diversity gain} of a code as defined above relates to the minimum rank of $(X-X')$ over all pairs of distinct code matrices $(X,X') \in \mc{X}^2$. Thus, the minimum rank of $\mc{X}$ should satisfy  
\begin{align*}
	\min_{X \neq X'} \rk{X-X'} = \min\{n_t,T\}=n_t. 
\end{align*}
A code satisfying this criterion is called a \emph{full-diversity} code. 

On the other hand, the \emph{coding gain} is proportional to the determinant $\det\left((X-X')(X-X')^\dagger\right)$. As a consequence, the minimum taken over all pairs of distinct codewords,
\begin{align*}
		\min_{X \neq X'} \det\left((X-X')(X-X')^{\dagger}\right), 
\end{align*}
should be as large as possible. For the infinite code 
\begin{align*}
	\mc{X}_\infty = \left\{\left.\sum\limits_{i=1}^{k}{s_i B_i} \right| s_i \in \Z \right\},
\end{align*}
we define the \emph{minimum determinant} as the infimum
\begin{align*}
	\Delta_{\min}(\mc{X}_\infty) := \inf_{X \neq X'} \det\left((X-X')(X-X')^{\dagger}\right). 
\end{align*}
If $\Delta_{\min}(\mc{X}_\infty) > 0$, \emph{i.e.}, the determinants do not vanish as the code size increases, the code is said to have the \emph{non-vanishing determinant} property. 

We assume henceforth that the number of transmit antennas and channel delay coincide, $n_t = T =: n$. Given a lattice $\Lambda \subset \mat(n,\C)$, we have by linearity
\begin{align*}
	\Delta_{\min}(\Lambda) = \inf\limits_{0\neq X \in \Lambda}{\left|\det(X)\right|^2}.
\end{align*}
This implies that any lattice $\Lambda$ with non-vanishing determinants can be scaled so that $\Delta_{\min}(\Lambda)$ achieves any wanted nonzero value. Consequently, the comparison of two different lattices requires some sort of normalisation. Let $\Lambda$ be a full lattice with volume $\vl{\Lambda}$.
The \emph{normalised minimum determinant} and \emph{normalised density} of $\Lambda$ are the normalised quantities
\begin{align*}
	\delta(\Lambda) = \frac{\Delta_{\min}(\Lambda)}{\vl{\Lambda}^{\frac{1}{2n}}}; \quad 
	\eta(\Lambda) = \frac{\Delta_{\min}(\Lambda)^{2n}}{\vl{\Lambda}},
\end{align*}
and satisfy the relation $\delta(\Lambda)^2 = \eta(\Lambda)^{\frac{1}{n}}$. Thus, for a fixed minimum determinant, the coding gain can be increased by maximising the density of the code lattice. Or, the other way around, for a fixed volume, the coding gain can be increased by maximising the minimum determinant of the lattice.

\subsubsection{Constructions from Cyclic Division Algebras}
\label{subsubsec:code_construction}

We move on to illustrate how space--time codes satisfying the two introduced criteria can be designed. We begin by ensuring full diversity, to which end the following result is helpful.  
\begin{theorem}\cite[Prop.~1]{sethuraman:stc}
	Let $K$ be a field and $\mc{D}$ an index $n$ division $K$-algebra over a maximal subfield $L$. Any finite subset $\mc{X}$ of the image of a ring homomorphism $\phi: \mc{D} \mapsto \mat(n, L)$ satisfies $\rk{X-X'} = n$ for any distinct $X, X' \in \mc{X}$. 
\end{theorem}	
	This leads to a straightforward approach for constructing full-diversity codes, namely by choosing the underlying structure to be a division algebra. In the same article, cyclic division algebras were proposed for code construction as a particular example of division algebras. The ring homomorphism $\phi$ is the link between the division algebra and a full-diversity space--time code, as we illustrate in the following. 

Let $\mc{C} = (L/K,\sigma,\gamma)$ be a cyclic division algebra of degree $n$. The left-regular representation $\rho: \mc{C} \to \mat(n,\C)$ is an injective ring homomorphism. We identify elements in $\mc{C}$ with elements in $\mat(n,\C)$ via $\rho$. This leads to the following definition.

\begin{definition}
 	Let $\mc{C}$ be a cyclic division algebra of index $n$ with left-regular representation $\rho: \mc{C} \to \mat(n,\C)$. A \emph{space--time code} constructed from $\mc{C}$ is a finite subset 
 	\begin{align*}
 		\mc{X} \subset \rho(\mc{C}).
 	\end{align*}
\end{definition}
To be consistent with Definition~\ref{def:stc}, let $\left\{ B_i \right\}_{i=1}^{k}\subset \mat(n,\C)$ with $k\leq 2n^2$ be a matrix basis for $\mc{C}$ over $\Q$. For a fixed signalling alphabet $S \subset \Z$, symmetric around the origin, the space--time code $\mc{X}$ is of the form
\begin{align*}
	\mc{X} = \left\{\left. \sum\limits_{i=1}^{k}{s_i B_i} \right| s_i \in S \right\}. 
\end{align*}

Note that, given an element $X = \rho(x)$, where $x \in \mc{C}$, we have that $\det(X) = \det(\rho(x)) \in K$. We can however restrict the entries of the codewords to certain subrings of the cyclic division algebra, for instance an order $\Gamma$.  For any $x \in \Gamma$, we have $\det(\rho(x)) \in \mc{O}_K$. This yields  $\det(\rho(x)) \ge 1$ for $K = \Q$ or $K$ an imaginary quadratic number field. Then, we can consider finite subsets of $\rho(\Gamma)$ as space--time lattice codes guaranteeing non-vanishing determinants (cf. Remark \ref{rednorm}). 

\begin{example}
	Consider a MIMO system with $n=n_t = T = 2$, and consider the index-$2$ number field extension $L/K = \Q(i,\sqrt{5})/\Q(i)$. The ring of integers of $L$ is $\mc{O}_L = \Z[i,\theta]$ with $\theta = \frac{1+\sqrt{5}}{2}$, and we pick the relative integral basis $\left\{1,\theta\right\}$ of $\mc{O}_L$ over $\mc{O}_K=\Z[i]$. The \emph{Golden code} \cite{belfiore:golden} is constructed from the cyclic division algebra 
	\begin{align*}
		\mc{G} = (L/K,\sigma,\gamma) \cong (5,\gamma)_{\Q(i)}
	\end{align*}
	with $\sigma: \sqrt{5} \mapsto -\sqrt{5}$ and $\gamma \in \Q(i)$ non-zero and such that $\gamma \neq \nm{L/K}{l}$ for any $l \in L$. We pick $\gamma = i$, leading to a matrix representation of  $\mc{G}$  of the form 
	\begin{align*}
		X = \rho(x) = \begin{bmatrix} x_0+\theta x_1 & i(x_2+\sigma(\theta) x_3) \\ x_2+\theta x_3 & x_0+\sigma(\theta)x_1 \end{bmatrix},
	\end{align*}	
where $x_i\in K$.	
	
	The algebra $\mc{G}$ is a division algebra by Theorem~\ref{thm:non_norm}, so that the Golden code is indeed a full-diversity space--time code. Moreover, by restricting the codewords to the natural  order $\Gamma_{nat}$ by choosing  $x_i \in \mc{O}_K$ guarantees the non-vanishing determinant property (cf. Remark \ref{rednorm}). 
	
	The actual Golden code lattice is a twisted version of $\rho(\Gamma_{nat})$ in order to get an orthogonal lattice. The twisting does not affect the normalised minimum determinant. 
\end{example}
\newpage

\section{Codes with Reduced ML Decoding Complexity}
\label{sec:reduced_ml}

Using multiple antennas for increased diversity --- and additionally enabling temporal diversity --- comes at the cost of a higher complexity in decoding. The worst-case complexity of maximum-likelihood (ML) decoding is upper bounded by that of exhaustive search, and is often computationally too expensive for practical use for higher-dimensional code lattices. A fast-decodable space--time code is, in colloquial terms, simply a space--time code whose worst-case ML decoding complexity is lower than that of exhaustive search. 

Yet, independently of the actual decoder used, the ML decoding complexity of a space--time code can sometimes be reduced by algebraic means, allowing for parallelisation in the ML search.  If the underlying code lattice is of rank $k$, this requires in principle joint decoding of $k$ information symbols. One way to achieve fast-decodability (this is also how we define fast decodability more formally below) is then to reduce the dimensionality of the (\emph{e.g.}, sphere) decoder, that is, to enable parallelisation where each parallel set contains less than $k$ symbols to be jointly decoded. 

In this section we introduce the technique of ML decoding and revise criteria for a space--time code to be fast-decodable. We further specify different families of fast-decodable codes and study their potential decoding complexity reduction.

\subsection{Maximum-Likelihood Decoding}
\label{subsec:ml}

In the previous sections we have seen what properties a space--time code should exhibit to potentially ensure a reasonable performance, at least in terms of reliability. There are however more aspects of the communication process which need to be taken into consideration. Orthogonal lattices allow for efficient encoding of the information symbols and bit-labelling of the codewords, while not necessarily yielding the best possible error performance. On the other hand, a too complicated lattice structure makes it more complex to encode a signal in the first place, and may require brute force bit labelling of the codewords. 

On the receiver's side, the structure of the code lattice determines the complexity of the decoding process. Indeed, as already mentioned, the major bottle-neck in effective implementation of algebraic space--time codes has been their decoding complexity. The concept of fast-decodability was introduced in \cite{biglieri:fd} in order to address the possibility for reducing the dimensionality of the ML decoding problem (cf. \eqref{eqn:ml_decod}) without having to resort to suboptimal decoding methods.

Given a finite signalling alphabet $S \subset \Z$, the ML decoding complexity of a rank-$k$ space--time code $\mc{X}$ is defined as the minimum number of values that have to be computed for finding the solution to \eqref{eqn:ml_decod}. The upper bound is the worst-case decoding complexity that we denote by $\mf{D}(S)$, which for its part is upper bounded by the exhaustive search complexity,  $\mf{D}(S)\le |S|^k$. The following definition is hence straightforward. 	
\begin{definition}
\label{def:fd_exp}
	A space--time code $\mc{X}$ is said to be \emph{fast-decodable} if its ML decoding complexity is upper bounded by
	\begin{align*}
		\mf{D}(S) = c|S|^{k'}, 
	\end{align*}
	where $k' < k-2$ and $c\leq k$ is a constant describing the number of parallel symbol groups to be decoded. If $c=k$ (and hence $k'=1$), this means that we can decode symbol-wise with linear complexity.  We refer to $k'$ as the \emph{complexity order}. 
\end{definition}

\begin{remark}
	We require $k' < k-2$ as we are considering the reduction of real dimensions. A reduction of 2 can be obtained merely by Gram-Schmidt orthogonalisation. 
\end{remark}

We proceed to investigate how to determine the complexity order of a space--time code $\mc{X}$. Let $\left\{B_i\right\}_{i=1}^k$ be a basis of $\mc{X}$ over $\Z$, and $X \in \mc{X}$ the transmitted signal. Recall the isometry \eqref{eqn:isometry},
which allows us to identify the space--time code lattice with a lattice in Euclidean space. In addition, for $c \in \C$ let 
\begin{align*}
	\tilde{c} = \begin{bmatrix} \Re(c) & -\Im(c) \\ \Im(c) & \Re(c) \end{bmatrix}.
\end{align*}

From the channel output $Y = HX + N$, define the matrices
\begin{align*}
	B &= \begin{bmatrix} \iota(B_1) & \cdots & \iota(B_k) \end{bmatrix} \in \mat(2n_t T \times k,\R), \\
	B_H &= \begin{bmatrix} \iota(HB_1) & \cdots & \iota(HB_k) \end{bmatrix} \in \mat(2n_r T\times k,\R).
\end{align*} 
The equivalent received codeword under the isometry can be expressed as $\iota(HX) = B_H\mb{s}$ for a coefficient vector $\mb{s}^t = (s_1,\ldots,s_k) \in S^k$, and we get an equivalent vectorized channel equation 
\begin{align*}
	\iota(Y) &= B_H\mb{s} + \iota(N) \\
	&= (I_T \otimes \tilde{H})B \mb{s} + \iota(N),
\end{align*}
where $\tilde{H} = (\tilde{h}_{ij})_{i,j}$ and $\otimes$ denotes the Kronecker product. 

We go on to perform $QR$-decomposition on $B_H$, or equivalently on $(I_T\otimes \tilde{H})B$. We write $B_H = QR$ with $Q \in \mat(2n_r T \times k,\R)$ unitary and $R \in \mat(k,\R)$ upper triangular. More precisely, if we write 
\begin{align*}
	B_H = \begin{bmatrix} \mb{b}_1 & \cdots & \mb{b}_k \end{bmatrix}, \qquad 
	Q = \begin{bmatrix} \mb{q}_1 & \cdots & \mb{q}_k \end{bmatrix},
\end{align*}
the matrix $R$ is precisely given by
\begin{align*}
	R = \begin{bmatrix}
		\|\mb{r}_1\| & \langle \mb{q}_1,\mb{b}_2 \rangle & \langle \mb{q}_1,\mb{b}_3 \rangle & \cdots & \langle \mb{q}_1,\mb{b}_k \rangle \\ 0 & \|\mb{r}_2\| & \langle \mb{q}_2,\mb{b}_3 \rangle & \cdots & \langle \mb{q}_2,\mb{b}_k \rangle \\
		\vdots & & \ddots & & \vdots \\
		0 & 0 & \cdots & 0 & \|\mb{r}_k\|
	\end{bmatrix},
\end{align*}
where 
\begin{align*}
	\mb{r}_1 = \mb{b}_1; \quad \mb{r}_i = \mb{b}_i - \sum\limits_{j=1}^{i-1}\frac{\langle \mb{q}_j,\mb{b}_i\rangle}{\langle \mb{q}_j,\mb{q}_j\rangle} \mb{q}_j ;\quad \mb{q}_i = \frac{\mb{b}_i}{\|\mb{b}_u\|}.
\end{align*}

Since the receiver has channel state information, and as the noise is zero-mean, the decoding process, as we have already seen, requires to solve the minimisation problem 
\begin{align*}
	\hat{X} = \argmin\limits_{X \in \mb{X}}{\|Y-HX\|_F^2}.
\end{align*}

Using the $QR$ decomposition, we can solve the equivalent problem 
\begin{align*}
	\hat{\mb{s}} = \argmin\limits_{\mb{s} \in S^k}{\|\iota(Y) - B_H\mb{s}\|^2} = \argmin\limits_{\mb{s} \in S^k}{\|Q^\dagger \iota(Y) - R\mb{s}\|^2},
\end{align*}
a problem which can be solved using a real sphere-decoder \cite{viterbo:sphere_decod}. It is now clear that the structure of the matrix $R$ determines the complexity of decoding. With zero entries at specific places, the involved variables can be decoded independently of each other, allowing for parallelisation in the decoding process, and potentially reducing the decoding complexity. 

Moreover, different orderings of the weight matrices $B_i$, or equivalently of the symbols $s_i$, result in different zero patterns in the matrix $R$. An algorithm for the optimal ordering of the weight matrices  resulting in the minimum possible decoding complexity is given in \cite{jithamithra:sphere_decod}, and was implemented in \cite{jaameri:fd}. We use the implementation found in the latter article for the explicit computation of the decoding complexity reduction of the example codes exposed in the remaining of this section.
 
Before we move on to define more specialized families of fast-decodable codes, we present the usual approach to give sufficient conditions for a code to be fast-decodable. This so-called \emph{Hurwitz-Radon quadratic form} approach is discussed in \cite{srinath:low_ml, jithamithra:quadratic, jithamithra:sphere_decod}, among others. The idea behind the Hurwitz Radon approach on which the quadratic form is based is to give a criterion for when two variables of the considered code can be decoded independently. More specifically, the variables $s_i, s_j$ can be decoded independently if their corresponding weight matrices $B_i$, $B_j$ are \emph{mutually orthogonal}, \emph{i.e.}, 
\begin{align*}
	B_i B_j^\dagger + B_j B_i^\dagger = 0.
\end{align*}

To be more precise, we give the following result
\begin{proposition}\cite[Thm.~2]{srinath:low_ml}
	Let $\mc{X}$ be a space--time code of rank $k$ with weight matrices $\left\{B_i\right\}_{i=1}^{k}$. If the matrices $B_i$ and $B_j$ are mutually orthogonal, then the columns $\mb{b}_{i}$ and $\mb{b}_{j}$ of $B_H$ are orthogonal. 
\end{proposition}	

In particular, the entry $(i,j)$ of the associated matrix $R$ is zero. Relating to this condition, the Hurwitz-Radon quadratic form is a tool which allows to deduce the actual ML decoding complexity of a space--time code based on the mutually orthogonality of the weight matrices. In particular, the criterion based on the quadratic form shows that fast decodability can be achieved solely by designing the weight matrices cleverly, and is independent of the channel and number of antennas. We give the following definition.
\begin{definition}
	Let $\mc{X}$ be a space--time code of rank $k$, and let $X \in \mc{X}$. The \emph{Hurwitz-Radon quadratic form} is the map
	\begin{align*}
		\mc{Q}: \mc{X} &\to \R, \\
		X = \sum\limits_{i=1}^k{B_i s_i} &\mapsto \sum\limits_{1 \le i \le j \le k}{s_i s_j \delta_{ij}},
	\end{align*}
where $\delta_{ij} := \|B_i B_j^\dagger + B_j B_i^\dagger\|_F^2$.
\end{definition}

Note that $B_i, B_j$ are mutually orthogonal if and only if $\delta_{ij} = 0$.

\subsubsection{Multi-Group Decodable Codes}
\label{subsubsec:multigroup}

We begin with the family of multi-group decodable codes. 

\begin{definition}
	Consider a space--time code $\mc{X}$ defined by the weight matrices $\left\{B_i\right\}_{i=1}^k$.
	\begin{itemize}
		\item[i)] The code is \emph{$g$-group decodable} if there exists a partition of $\left\{1,\ldots,k\right\}$ into $g$ non-empty subsets $\Gamma_1,\ldots,\Gamma_k$ such that for $i \in \Gamma_u$, $j \in \Gamma_v$ with $u \neq v$, the matrices $B_i$ and $B_j$ are mutually orthogonal.  
		
		\item[ii)] The code is \emph{conditionally $g$-group decodable} if there exists a partition of $\left\{1,\ldots,k\right\}$ into $g+1$ non-empty subsets $\Gamma_1,\ldots,\Gamma_g, \Gamma$ such that for $i \in \Gamma_u, j \in \Gamma_v$ with $1 \le u < v \le g$,  the matrices $B_i$ and $B_j$ are mutually orthogonal.
	\end{itemize} 
\end{definition}

The family of codes which we refer to as conditionally $g$-group decodable codes are in literature also known as \emph{fast ML decodable codes}. We use the terminology of conditionally $g$-group decodable to not confuse the general notion of fast decodability with this specific family of fast-decodable codes. 

In the following, we consider a space--time code $\mc{X}$ with weight matrices $\left\{B_i\right\}_{i=1}^k$ and corresponding real information symbols $s_1,\ldots,s_k\in S$. For $\mc{X}$ $g$-group decodable or conditionally $g$-group decodable, we may without loss of generality order the variables according to the $g$ groups $\Gamma_1,\ldots,\Gamma_g$, \emph{i.e.}, 
	\begin{align}
	\label{eqn:ordering}
	\begin{split}
		&\left\{s_1,\ldots,s_{|\Gamma_1|}\right\} \in \Gamma_1, \\
		&\left\{s_{|\Gamma_1|+1},\ldots,s_{|\Gamma_1|+|\Gamma_2|}\right\} \in \Gamma_2, \quad \\
		&\qquad\qquad\qquad\vdots \\
		&\left\{s_{\sum\limits_{i=1}^{g-1}{|\Gamma_i|}+1},\ldots,s_{\sum\limits_{i=1}^{g-1}{|\Gamma_i|}+|\Gamma_g|}\right\} \in \Gamma_g.
	\end{split}
	\end{align}
We have the following result. 	
\begin{proposition}\cite[Lemma~1]{jithamithra:quadratic}
Let $\mc{X}$ be a $g$-group decodable space--time code, and let $M = (m_{ij})_{i,j}$ be the Hurwitz-Radon quadratic form matrix and $R = (r_{ij})_{i,j}$ the $R$-matrix from the $QR$ decomposition of $B_H$. Then, $m_{ij} = r_{ij} = 0$ for $i < j$ whenever $s_i \in \Gamma_u$ and $s_j \in \Gamma_v$ with $u \neq v$. In particular, the $R$-matrix takes the form 
\begin{align*}
	R = \begin{bmatrix} D_1 & & \\ & \ddots & \\ & & D_g \end{bmatrix},
\end{align*} 
where $D_i \in \mat(|\Gamma_i|,\R)$ is upper triangular, $1 \le i \le g$, and the empty spaces are filled with zeros. 
\end{proposition}

\begin{example}
\label{exp:alamouti}
	The first example we give is the complexity order of the Alamouti code $\mc{X}_A$ (cf. Section~\ref{sec:algebra}).
	We recall that this code consists of codewords
	\begin{align*}
		X = \begin{bmatrix} 
			x_1 + i x_2 & -(x_3 - i x_4) \\ x_3 + ix_4 & x_1 - i x_2
		\end{bmatrix},
	\end{align*}
	where $(x_1,x_2,x_3,x_4) \in \Z^4$ are usually taken to be integers to guarantee non-vanishing determinants. 
	
	The $R$-matrix associated with this code is in fact a diagonal $4\times 4$ matrix with equal diagonal entries. Hence, $\mc{X}_A$ is $4$-group decodable, and exhibits a complexity order $k'=1$. In other words, it is single-symbol decodable. 
\end{example}

\begin{example}
	We recall the code constructed in \cite[Ex.~6]{barreal:fd_stc_relay}. Consider the cyclic division algebra 
	\begin{align*}
		\mc{C}	= \left(F(\sqrt{-3},i)/F(i),\sigma,-\frac{2}{\sqrt{5}}\right),
	\end{align*}
	where $F = \Q(\sqrt{5})$ and $\sigma:\sqrt{-3} \mapsto -\sqrt{-3}$ but fixes $i$. Let $\tau$ be a generator of the cyclic Galois group $\gal(F(i)/F)$, \emph{i.e.}, $\tau(i) = -i$. Consider codewords of the form 
	\begin{align*}
		X = \begin{bmatrix}
			X_1 & \tau(X_1) \\
			X_2 & \tau(X_2)
		\end{bmatrix},
	\end{align*}
	where $\tau$ acts element-wise, and for $\theta = \frac{1+\sqrt{-3}}{2}$ and $k = 1,2$ we have 
	\begin{align*}
		X_k = \begin{bmatrix}
			x_{k,1}+x_{k,2}\theta & -\sqrt{-\gamma}(x_{k,3}+x_{k,4}\sigma(\theta)) \\ 
			\sqrt{-\gamma}(x_{k,3}+x_{k,4}\theta) & x_{k,1}+x_{k,2}\sigma(\theta)
		\end{bmatrix}
	\end{align*}
	with $x_{k,j}\in\mc{O}_{F(i)}$. With some abuse of notation, we have that $\tau(\sqrt{-\gamma})=\sqrt{-\gamma}$. Hence, each $X_k$ corresponds to the left-regular representation of an element in the natural order $\Gamma_{nat}\subseteq \mc{C}$, after balancing the effect of $\gamma$ by spreading it on the diagonal\footnote{This is a usual trick to balance the average energies of the codeword entries more evenly.  See \cite[Ex.~1]{barreal:fd_stc_relay} for more details.}. 
	
	The complexity of exhaustive search for a signalling alphabet $S$ is $|S|^{32}$. The above code, however, is $2$-group decodable. In fact, the associated $R$-matrix is of the form  
	\begin{align*}
		R = \begin{bmatrix} D_1 & \\ & D_2 \end{bmatrix}
	\end{align*}
	with $D_i \in \mat(16,\R)$ upper triangular. The code hence exhibits a complexity order  $k'=16$, resulting in a reduction of $50\%$.
\end{example}

In the case of conditionally $g$-group decodable codes, \emph{i.e.}, where we have a further non-empty group $\Gamma$, the $R$ matrix is not entirely block-diagonal. Instead, we have the following result.
\begin{proposition}\cite[Lem.~2]{berhuy:fd}
\label{prop:gd}
Let $\mc{X}$ be a conditionally $g$-group decodable code, and let $M = (m_{ij})_{i,j}$ be the Hurwitz-Radon quadratic form matrix and $R = (r_{ij})_{i,j}$ the $R$-matrix from the $QR$ decomposition. Then, $m_{ij} = r_{ij} = 0$ for $i < j$ whenever $s_i \in \Gamma_u$ and $s_j \in \Gamma_v$ with $1 \le u < v \le g$. In particular, the $R$-matrix takes the form
\begin{align*}
	R = \begin{bmatrix} D_1 & & & N_1 \\ & \ddots & & \vdots \\ & & D_g & N_g \\ & & & N \end{bmatrix},
\end{align*}
with $D_i \in \mat(|\Gamma_i|,\R)$ and $N \in \mat(|\Gamma|,\R)$ are upper triangular, and $N_i \in \mat(|\Gamma_i|\times|\Gamma|,\R)$. 
\end{proposition}

\begin{example}
	As an example of a conditionally $g$-group space--time code we recall the famous \emph{Silver code} \cite{paredes:silver, hollanti:silver}. The code is contained as a subset in the cyclic division algebra
	\begin{align*}
		\mc{C} = (\Q(i,\sqrt{-7})/\Q(\sqrt{-7}),\sigma,\gamma),
	\end{align*}
	Note that $\sigma$ is not just complex conjugation, as $\sigma(i) = -i$ and $\sigma(\sqrt{7}) = -\sqrt{7}$. With $\gamma = -1$, the algebra is division, and the resulting code is fully diverse and has non-vanishing determinants. The Silver code is however not directly constructed as a subset of $\rho(\Gamma)$ for $\Gamma$ an order of $\mc{C}$. Instead, it is defined as 
	\begin{align*}
		\mc{X}_{S} = \left\{\left.X = X_A(x_1,x_2) + TX_B(x_3,x_4) \right| x_1,\ldots,x_4 \in \Z[i]\right\},
	\end{align*}
	where $x_1,\ldots,x_4 \in \Z[i]$ and
	\begin{align*}
		T &= \begin{bmatrix} 1 & 0 \\ 0 & -1 \end{bmatrix} ;\qquad
		X_A(x_1,x_2) = \begin{bmatrix} x_1 & -x_2^\ast \\ x_2 & x_1^\ast \end{bmatrix}; \\
		X_B(x_3,x_4) &= \frac{1}{\sqrt{7}} \begin{bmatrix} (1+i)x_3 + (-2+2i)x_4 & -(1-2i)x_3^\ast-(1+i)x_4^\ast \\ (1+2i)x_3 + (1-i)x_4 & (1-i)x_3^\ast+(-1-2i)x_4^\ast \end{bmatrix}.
	\end{align*}
	
	In particular, a codeword is of the form 
	\begin{equation*}
	\resizebox{\textwidth}{!}{$
		X = \frac{1}{\sqrt{7}} \begin{bmatrix} x_1\sqrt{7}+(1+i)x_3+(-1+2i)x_4 & 
		-x_2^\ast\sqrt{7}-(1-2i)x_3^\ast-(1+i)x_4^\ast \\
		x_2\sqrt{7}-(1+2i)x_3-(1-i)x_4 & 
		x_1^\ast\sqrt{7}-(1-i)x_3^\ast-(-1-2i)x_4^\ast	
		\end{bmatrix}.$}
	\end{equation*}
	
Using the optimal ordering of the weight matrices, we find that the complexity order of the Silver code is $k' = 5$, resulting in a complexity reduction of $37.5\%$. 
\end{example}

\begin{example}
	As a second example, we recall the Srinath-Rajan code, originally proposed in \cite{srinath:low_ml} for a $4\times 2$-MIMO channel. To the best of the author's knowledge, this is the best performing code known for a $4\times 2$ system among codes with the same complexity order. We recall the construction illustrated in \cite{srinath:fd_nvd}, where the underlying algebraic structure was discovered. 
	
	Let $L/F$ be a cyclic Galois extension with cyclic Galois group $\gal(L/F) = \langle \tau \rangle$, and consider a cyclic division algebra $\mc{C}' = (L/F,\tau,\gamma')$. Moreover, let $\mc{C} = (L/K,\sigma,\gamma)$ be a cyclic division algebra of degree $n$, where $K \neq F$ and $\tau\sigma = \sigma\tau$. We require $\gamma \in K\cap F$ and $\gamma' \in F\backslash K$.
	
	For the $4\times 2$ Srinath-Rajan code, we make the choices 
	\begin{itemize}
		\item[i)] $L = \Q(i,\sqrt{5})$, $K = \Q(\sqrt{5})$, $F = \Q(i)$.
		\item[ii)] $\mc{C}' = (L/F,\tau,\gamma')$ with $\gamma' = i \notin K$ and $\tau:\sqrt{5} \mapsto -\sqrt{5}$. This cyclic division algebra gives rise to the Golden code. 
		\item[iii)] $\mc{C} = (L/K,\sigma,\gamma)$ with $\gamma = -1$ and $\sigma: i \mapsto -i$.  
	\end{itemize}
	
	Fix the $F$-basis $\left\{\theta_1,\theta_2\right\}$ of $L$, with $\theta_1 = 1+i(1-\theta), \theta_2 = \theta_1\theta \in \mc{O}_L$, where $\theta = \frac{1+\sqrt{5}}{2}$. Codewords are of the form 
	\begin{align*}
		X = \begin{bmatrix} 
			x_0 & -\sigma(x_1) & i\tau(x_2) & -i\tau\sigma(x_3) \\
			x_1 & \sigma(x_0) & i\tau(x_3) & i\tau\sigma(x_2) \\
			x_2 & -\sigma(x_3) & \tau(x_0) & -\tau\sigma(x_1) \\
			x_3 & \sigma(x_2) & \tau(x_1) & \tau\sigma(x_0)		
		\end{bmatrix},
	\end{align*}
where $x_i = x_{i1}\theta_1 + x_{i2}\theta_2$ with $x_{ij} \in \Z[i]$. 
	
	This code is conditionally $4$-group decodable, where 8 real variables need to be conditioned, and the remaining 8 variables can be grouped in 4 groups of 2. This can be seen from the structure of the $R$-matrix, which for this code takes the form 
	\begin{align*}
		R = \begin{bmatrix} D_1 & & & & N_1 \\  & D_2 & & & N_2 \\ & & D_3 & & N_3 \\ & & & D_4 & N_4 \\ & & & & N \end{bmatrix},	
	\end{align*}
where $D_i$ are $2\times 2$ upper triangular matrices, $N_i$ are $2\times 8$ matrices, and $N$ is an $8\times 8$ upper triangular matrix. 
This yields a decoding complexity order $k' = 10$. This is a reduction in complexity of $37.5\%$. 
\end{example}

To summarize, we observe that the $R$ matrix allows to directly read the decoding complexity of a $g$-group decodable and conditionally $g$-group decodable code. After conditioning the last $|\Gamma|$ variables, the variables in each group $\Gamma_i$ can be decoded independently of the other groups. This is summarized in the following result. 

\begin{theorem}
	The decoding complexity order of a (conditionally) $g$-group decodable code $\mc{X}$ with possibly empty subset $\Gamma$ is given by 
	\begin{align*}
		k' = {|\Gamma|+\max\limits_{1 \le i \le g}{|\Gamma_i|}}.
	\end{align*}
\end{theorem}

Unfortunately, there is a trade-off between the maximum rate and maximum decoding complexity reduction of space--time codes. The recent work \cite{berhuy:fd_bounds} treats both these questions for multi-group decodable codes by analysing the mutually orthogonality of matrices in central simple subalgebras of $\mat(n,\C)$ over number fields. The authors show on one hand that there is a lower bound for the decoding complexity of full-rate $n\times n$ space--time codes. They furthermore derive an upper bound on the number of groups of a multi-group decodable code. We summarise the results in the following theorem.

\begin{theorem}\cite{berhuy:fd_bounds}
	Let $\mc{X}$ be an $n\times n$ space--time code defined by the weight matrices $\left\{B_i\right\}_{i=1}^{2k^2}$, and let $S$ denote the employed real signalling alphabet. 
	\begin{itemize}
		\item[i)] If $\mc{X}$ is full-rate, then the decoding complexity order is not better than ${n^2+1}$. 
		
		\item[ii)] If $\mc{X}$ is multi-group decodable, the number of groups $g$ is upper bounded by 
		\begin{align*}
			g \le 2\nu_2(n)+4,
		\end{align*}
		where $\nu_2(n)$ denotes the highest power of $2$ dividing $n$. In particular, if the weight matrices are chosen from a $K$-central division algebra with $K$ a number field, we have $g \le 4$. 
	\end{itemize}
\end{theorem}

\subsubsection{Fast-Group Decodable Codes}
\label{subsubsec:fastgroup}

Fast-group decodable codes combine the structure of the block-diagonal $R$-matrix with further parallelisation within each of the independent groups. We start with the formal definition. 
\begin{definition}
	Consider a space--time code $\mc{X}$ defined by the weight matrices $\left\{B_i\right\}_{i=1}^k$. The code is \emph{fast-group decodable} if 
	\begin{itemize}
		\item[a)] There is a partition of $\left\{1,\ldots,k\right\}$ into $g$ non-empty subsets $\Gamma_1, \ldots, \Gamma_g$ such that whenever $i \in \Gamma_u$, $j \in \Gamma_v$ with $u \neq v$, the matrices $B_i$ and $B_j$ are mutually orthogonal. 
		
		\item[b)] In addition, for at least one group $\Gamma_i$, we have $\langle \mb{q}_{l_1},\mb{b}_{l_2}\rangle = 0$, where $l_1 = 1,\ldots L_i -1$ and $l_2 = l_1+1,\ldots, L_i$ with $L_i \le |\Gamma_i|$. 
	\end{itemize}
\end{definition}

Consider a fast-group decodable space--time code $\mc{X}$, and denote by $\Gamma_1,\ldots,\Gamma_g$ the groups in which the corresponding symbols can be jointly decoded. Assume that the variables $s_1,\ldots,s_k$ are without loss of generality ordered according to their groups, as described above \eqref{eqn:ordering}.
	
\begin{proposition}\cite[Lem.~3]{jithamithra:quadratic}
\label{prop:fast_gd}
Let $\mc{X}$ be a $g$ fast-group decodable space--time code, and let $M = (m_{ij})_{i,j}$ be the Hurwitz-Radon quadratic form matrix and $R = (r_{ij})_{i,j}$ the $R$-matrix from the $QR$ decomposition. Then, $m_{ij} = r_{ij} = 0$ for $i < j$ whenever $s_i \in \Gamma_u$, $s_j \in \Gamma_v$ with $u \neq v$. Furthermore, each group $\Gamma_i$ admits to remove $L_i$ levels from the sphere-decoder tree if $m_{i_{l_1} i_{l_2}} = 0$, where $l_1 = 1,\ldots,L_i-1$ and $l_2 = l_1+1,\ldots,L_i$, with $L_i \ge |\Gamma_i|$. In particular, the $R$-matrix takes the form 
\begin{align*}
	R = \begin{bmatrix} R_1 & & \\ & \ddots & \\ & & R_g \end{bmatrix},
\end{align*} 
where the empty spaces are filled with zeros. Each of the matrices $R_i \in \mat(|\Gamma_i|,\R)$ is of the form 
\begin{align*}
	R_i = \begin{bmatrix} D_i & B_{i_1} \\ & B_{i_2} \end{bmatrix},
\end{align*}
with $D_i \in \mat(L_i,\R)$ is upper triangular, $B_{i_2}$ is a square upper triangular matrix and $B_{i_1}$ is a rectangular matrix.
\end{proposition}

\begin{theorem}
	The decoding complexity of a $g$ fast-group decodable space--time code $\mc{X}$ with real signalling alphabet $S$ is given by 
	\begin{align*}
		\mf{D}(S) = |S|^{\max\limits_{1 \le i \le g}{|\Gamma_i|}-L},
	\end{align*}
	where $L$ corresponds to the number of levels that can be removed from the sphere-decoder tree for jointly decoding $\max\limits_{1 \le i \le g}{|\Gamma_i|}$ symbols.
\end{theorem}

\begin{example}
	The authors in \cite{ren:group_decod} construct a $4\times 4$ fast-group decodable code based on an orthogonal space--time code. Codewords are of the form 
	\begin{equation*}
	\resizebox{\textwidth}{!}{$
		X = \begin{bmatrix}
		x_1+ix_2+ix_{15}+ix_{16}+ix_{17} & x_7+ix_8+x_{13}+ix_{14} & x_3+ix_4+x_{11}+ix_{12} & -x_5-ix_6+x_9+ix_{10} \\
		-x_7+ix_8-x_{13}+ix_{14} & x_1+ix_2+ix_{15}-ix_{16}-ix_{17} & x_5-ix_6+x_9-ix_{10} & x_3-ix_4-x_{11}+ix_{12} \\
		-x_3+ix_4-x_{11}+ix_{12} & -x_5-ix_6-x_9-ix_{10} & x_1-ix_2+ix_{15}-ix_{16}+ix_{17} & x_7-ix_8-x_{13}+ix_{14} \\
		x_5-ix_6-x_9+ix_{10} & -x_3-ix_4+x_{11}+ix_{12} & -x_7-ix_8+x_{13}+ix_{14} & x_1-ix_2+ix_{15}+ix_{16}-ix_{17} 
		\end{bmatrix},$}
	\end{equation*}
where $x_i$ are real symbols. We refer to the original paper for more details on the explicit construction. 
The algebraic structure of this code allows to remove $5$ levels from the sphere decoding tree. In particular, the decoding complexity order is $k' = 12$, resulting in a reduction in decoding complexity of $\sim 30\%$. 
\end{example}

\subsubsection{Block Orthogonal Codes}
\label{subsubsec:block_orth}

The last family of fast-decodable codes that we treat are \emph{block orthogonal codes}. We define this family by means of the structure of the associated $R$-matrix. 

\begin{definition}
	Let $\mc{X}$ be a space--time code. The code is said to be \emph{$g$-block orthogonal} if the associated $R$-matrix has the structure
	\begin{align*}
		R = \begin{bmatrix}
			R_1 & B_{12} & \cdots & B_{1g} \\  & R_2 & \cdots & B_{2g} \\
			 & & \ddots & \vdots \\ & & & R_g 
		\end{bmatrix},	
	\end{align*}
	where the empty spaces are filled with zeros and the matrices $B_{ij}$ are non-zero rectangular matrices. Further, the matrices $R_i$ are block diagonal matrices of the form 
	\begin{align*}
		R_i = \begin{bmatrix} 
			U_{i,1} & & \\ & \ddots & \\ & & U_{i,k_i} 
		\end{bmatrix},
	\end{align*}
	with each of the blocks $U_{i,j}$ is a square upper triangular matrix.
\end{definition}

Assuming that each of the matrices $R_i$ has the same number of blocks $k$, we can determine a block orthogonal code by the three parameters $(g,k,p)$, where $g$ is the number of matrices $R_i$, $k$ denotes the number of block matrices which compose each matrix $R_i$ and $p$ is the number of diagonal entries in the block matrices $U_{i,j}$. 

\begin{example}
	The aforementioned Golden code is a $(2,2,2)$ block orthogonal code. However, as its decoding complexity order is $k' = 6 < 8 = k$, it is not fast-decodable by the requirement of a strict inequality as per Definition~\ref{def:fd_exp}. 
	
	As an example of a fast-decodable block orthogonal code, we consider the $(2,4,2)$ block orthogonal code from \cite{jithamithra:reduced_complexity}. For a signalling vector $(s_1,\ldots,s_{16})$, a codeword is of the form 
	\begin{align*}
		X = X'(s_1,\ldots,s_8) + \begin{bmatrix} 
			1 & 0 & 0 & 0 \\ 0 & -1 & 0 & 0 \\ 0 & 0 & 1 & 0 \\ 0 & 0 & 0 & -1
		\end{bmatrix} X'(s_9,\ldots,s_{16}),
	\end{align*}
	where 
	\begin{equation*}
		\resizebox{\textwidth}{!}{$
		X'(s_1,\ldots,s_8) = \begin{bmatrix} 
			(s_1-s_2)+i(s_3-s_4) & 0 & (s_7-s_8)+i(s_5-s_6) & 0 \\
			0 & (s_1-s_2)+i(s_4-s_3) & 0 & (s_8-s_7)+i(s_6-s_5) \\
			-(s_7+s_8)+i(s_5+s_6) & 0 & (s_1+s_2)-i(s_3+s_4)& 0 \\ 0 & (s_7+s_8)-i(s_5+s_6) & 0 & (s_1+s_2)+i(s_3+s_4) 
		\end{bmatrix},$}
	\end{equation*}
\end{example}

\begin{remark}
Recall that the property of fast decodability relates to the reduction in decoding complexity without resorting to suboptimal decoding methods. By modifying the decoding algorithm used, the decoding complexity of certain codes can be lowered. For example, the main algorithm of \cite{sirinaunpiboon:golden_ml} reduces the complexity order of the Golden code from $k = 6$, corresponding to the complexity of ML-decoding, to $k' = 4$, while maintaining nearly-ML performance. The algorithm is specific to the Golden code, but has been generalized to the $3\times 3$ and $4\times 4$ perfect codes in, respectively, \cite{howard:ml_decod, amani:ml_decod}. 
\end{remark}

In contrast to the previously introduced families, the approach via the Hurwitz-Radon quadratic form does not capture the complexity reduction for block orthogonal codes. This was recently addressed in \cite{mejri:fd_revisited}, where relaxed conditions are derived for classifying codes into the here treated families of fast-decodable codes. More precisely, for block orthogonal codes we do not have an analogue of Proposition~\ref{prop:gd} or \ref{prop:fast_gd} relating the matrix $M$ of the quadratic form to the $R$-matrix in the $QR$ decomposition of $B_H$.

\subsection{Inheriting Fast Decodability}
\label{subsec:iterative}

Crucial for space--time codes to exhibit desirable properties is the underlying algebraic framework. Constructing codes for larger number of antennas means dealing with higher degree field extensions and algebras, which are harder to handle. We briefly recall an iterative space--time code construction proposed in \cite{markin:iterated} which, starting with an $n\times n$ space--time code, results in a new $2n\times 2n$ space--time code with the same code rate and double (lattice) rank. The advantage of this construction is that when applied carefully, the resulting codes inherit good properties from the original space--time codes. 

\begin{figure}[h!]
\centering
	\includegraphics[width=0.3\textwidth]{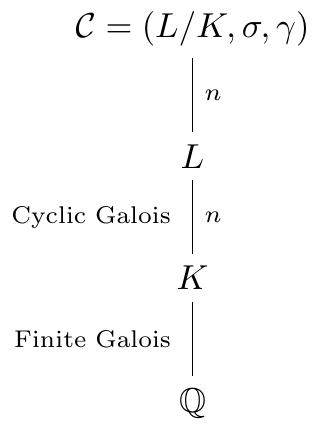}
	\caption{Tower of extensions for the MIMO example code.}
	\label{fig:iterated}
\end{figure}

As the general setup, consider the tower of extensions depicted in Figure~\ref{fig:iterated}. 

The cyclic Galois group of $L/K$ is generated by $\sigma$, \emph{i.e.}, $\gal(L/K) = \langle \sigma \rangle$, and we denote the left-regular representation by $\rho: \mc{C} \to \mat(n,L)$. Let $\tau \in \Aut(L)$ be an automorphism of $L$, and make the following assumptions:
\begin{align}
\label{eqn:iter_assumptions}
	\begin{split}
		\tau(\gamma) = \gamma; \quad \tau \sigma = \sigma \tau.
	\end{split}
\end{align}

By the above assumptions we have $\tau \rho = \rho \tau$, and it is clear that $\tau$ can be extended to an automorphism of $\mc{C}$ and $\rho(\mc{C})$, respectively, by 
\begin{align*}
	\tau\left(\sum\limits_{i=0}^{n-1}{e^i x_i}\right) = \sum\limits_{i=0}^{n-1}{e^i \tau(x_i)}; \qquad
	\tau\left((a_{ij})_{i,j}\right) = \left(\tau(a_{ij})\right)_{i,j}.
\end{align*}

We can now fix an element $\theta \in \mc{C}$, as well as a $\Q$-automorphism of $L$, $\tau \in \Aut_{\Q}(L)$, and have the following important definition.
\begin{definition}
\label{def:iterated_map}
Let $K$ be a finite Galois extension of $\Q$ and $\mc{C} = (L/K,\sigma,\gamma)$ be a cyclic division algebra of degree $n$. Fix $\theta \in \mc{C}$ and $\tau \in \Aut_{\Q}(L)$.
\begin{itemize}
	\item[(a)] Define the function
	\begin{align*}
			\alpha_{\tau,\theta}: \mat(n,L)\times\mat(n,L) &\to \mat(2n,L) \\
			(X,Y) &\mapsto \begin{bmatrix} X & \theta\tau(Y) \\ Y & \tau(X) \end{bmatrix}.
	\end{align*}
	
	\item[(b)] If $\theta = \zeta\theta'$ is totally real or totally imaginary, $\theta' > 0$ and $\zeta \in \left\{\pm 1, \pm i\right\}$, define the alike function	
	\begin{align*}
			\tilde{\alpha}_{\tau,\theta}: \mat(n,L)\times\mat(n,L) &\to \mat(2n,L) \\
			(X,Y) &\mapsto \begin{bmatrix} X & \zeta\sqrt{\theta'}\tau(Y) \\ \sqrt{\theta'}Y & \tau(X) \end{bmatrix}.
	\end{align*}
\end{itemize}	
\end{definition}

The defined maps restrict to $\mc{C}\times \mc{C} \to \mat(2,\mc{C})$ by identifying $x,y \in \mc{C}$ with their representation $X = \rho(x)$, $Y = \rho(y)$. 

Suppose that the algebra $\mc{C}$ gives rise to a rank-$k$ space--time code $\mc{X}$ defined via matrices $\left\{ B_i \right\}_{i=1}^{k}$. Then, the matrices $\left\{ \alpha_{\tau,\theta}(B_i,0), \alpha_{\tau,\theta}(0,B_i)\right\}_{i=1}^{k}$ (or applying $\tilde{\alpha}_{\tau,\theta}(\cdot,\cdot)$, respectively) define a rank-$2k$ code 
\begin{align*}
	\mc{X}_{\mathrm{it}} = \left\{\left. \sum\limits_{i=1}^{k}\left[s_i\alpha_{\tau,\theta}(B_i,0) + s_{k+i}\alpha_{\tau,\theta}(0,B_i)\right] \right| s_i \in S \right\}. 	
\end{align*}

We summarise the main results of \cite{markin:iterated} in the following proposition. 
\begin{proposition}
\label{prop:iterated_construction} \cite[Thm.~1, Thm.~2]{markin:iterated}
	Let $\mc{C} = (L/K, \sigma, \gamma)$ be a cyclic division algebra giving rise to a rank-$k$ space--time code $\mc{X}$ defined by the matrices $\left\{ B_i \right\}_{i=1}^{k}$. Assume that $\tau \in \Aut_{\Q}(L)$ commutes with $\sigma$ and complex conjugation, and further $\tau(\gamma) = \gamma$, $\tau^2 = \id$. Fix $\theta \in K^{\langle \tau \rangle}$, where $K^{\langle \tau \rangle}$ is the subfield of $K$ fixed by $\tau$. Identifying an element of $\mc{C}$ with its left-regular representation $\rho$, we have:
	\begin{itemize}
		\item[(i)] The image $\mc{I} = \alpha_{\tau,\theta}(\mc{C},\mc{C})$ is an algebra and is division if and only if $\theta \neq z\tau(z)$ for all $z \in \mc{C}$. Moreover, for any $\alpha_{\tau,\theta}(x,y) \in \mc{I}$, we have $\det(\alpha_{\tau,\theta}(x,y)) \in K^{\langle \tau \rangle}$. 
		
		\item[(ii)] If in addition $\theta = \zeta\theta'$ is totally real or totally imaginary, the image $\tilde{\mc{I}} = \tilde{\alpha}_{\theta}(\mc{C},\mc{C})$ retains both the full-diversity and non-vanishing determinant property. If for some $i,j$, $B_i B_j^{\dagger} + B_j B_i^{\dagger} = 0$, we have
		\begin{align*}
			\tilde{\alpha}_{\tau,\theta}(B_i,0)\tilde{\alpha}_{\tau,\theta}(B_j,0)^{\dagger} + \tilde{\alpha}_{\tau,\theta}(B_j,0)\tilde{\alpha}_{\tau,\theta}(B_i,0)^{\dagger} &= 0\,, \\
				\tilde{\alpha}_{\tau,\theta}(0,B_i)\tilde{\alpha}_{\tau,\theta}(0,B_j)^{\dagger} + \tilde{\alpha}_{\tau,\theta}(0,B_j)\tilde{\alpha}_{\tau,\theta}(0,B_i)^{\dagger} &= 0.
		\end{align*}
	\end{itemize}
\end{proposition}

The second part of Proposition~\ref{prop:iterated_construction}, in particular, states that under appropriate conditions, fast decodability is inherited from the rank-$k$ space--time code $\mc{X}$ to the iterated code $\mc{X}_\mathrm{it}$.
\newpage

\section{Explicit Constructions}
\label{sec:codes}

All the notions and concepts introduced in the previous chapters lead to this last part. To conclude the chapter, we focus on explicit construction methods for fast-decodable space--time codes

Throughout this chapter, we have provided multiple examples of space--time codes with reduced ML decoding complexity. Such examples can sometimes be found by chance, but most often a clever design gives rise to infinite families of codes with reduced decoding complexity. In the following, we turn our attention to communication setups for which such general results are known. To the best of the authors' knowledge, the here presented constructions are the only general fast-decodable algebraic constructions found in literature.

\subsection{Asymmetric Space--Time Codes}
\label{subsec:asymmetric}

Above we have exemplified the $4\times 2$ Srinath-Rajan code, the best performing code for this channel among codes with the same complexity order. Here, we discuss a methodology for constructing well-performing fast-decodable space--time codes for the $4\times 2$ MIMO channel, offering a reduction in decoding complexity of up to $37.5\%$.

The motivation behind the following construction is the structure of the Alamouti code (cf. Example~\ref{exp:alamouti}). As we have seen, the decoding complexity of the Alamouti code equals the size of the employed real signaling alphabet, $\mf{D}(S) = |S|$. Motivated by this observation, it is of interest to study space--time codes which are subsets of the rings $\mat(k,\mathbb{H})$. This motivates the next result. 

\begin{theorem}\cite{vehkalahti:asymmetric}
\label{thm:injection}
	Let $\mc{C}$ be a cyclic division algebra of degree $n$, with center $K$ of signature $(r,s)$, $r + 2s = m$. There exists an injection 
	\begin{align*}
		\psi: \mc{C} \hookrightarrow \diag\left(\mat(n/2,\mathbb{H})^w \times \mat(n,\R)^{r-w} \times \mat(n,\C)^s \right),
	\end{align*}
	where each $n\times n$ block is mapped to the corresponding diagonal block of a matrix in $\mat(mn,\C)$. Here, $w$ is the number of places which ramify in $\mc{C}$. 
	
	In particular, $\mc{C}$ can be embedded into $\mat(n/2,\mathbb{H})$ if 
	\begin{itemize}
		\item[i)]	The center $K$ is totally real, \emph{i.e.}, $r = m$. 
		\item[ii)]	The infinite places of $K$ are ramified in $\mc{C}$. 
	\end{itemize}
\end{theorem}

The ramification assumptions of places in the considered algebra are rather technical, and the interested reader is referred to \cite{vehkalahti:asymmetric} for further details.  

While the above result guarantees the existence of an injection into $\mat(n/2,\mathbb{H})$ when the conditions are satisfied, it does not make the embedding explicit. This is achieved in the following result. 

\begin{theorem}\cite[Prop.~11.1]{vehkalahti:asymmetric}
\label{thm:embedding}
 	Let $\mc{C} = (L/\Q,\sigma,\gamma)$ be a cyclic division algebra satisfying the requirements from Theorem~\ref {thm:injection}. Given for $x  \in \mc{C}$ an element $X = \rho(x) \in \mc{X}$, where $\mc{X}$ is a space--time code arising from the algebra $\mc{C}$, we have an explicit map
 	\begin{align*}
 		\psi: \mc{C} &\to \mat(n_t/2,\mathbb{H}) \\
 		X &\mapsto BPX(BP)^{-1}, 
 	\end{align*}
 	where $P = (p_{ij})_{i,j}$ is a permutation matrix with entries 
 	\begin{align*}
 		p_{ij} = \begin{cases}
 			1 &\mbox{ if } 2\nmid i \text{ and } j = \frac{i+1}{2}, \\
 			1 &\mbox{ if } 2\mid i \text{ and } j = \frac{i+n_t}{2}, \\
 			0 &\mbox{ otherwise},
 		\end{cases}
 	\end{align*}
 	and $B = \diag(\sqrt{|\gamma|},|\gamma|,\ldots,\sqrt{|\gamma|},|\gamma|)$. 
\end{theorem}

We now turn our attention to the $4\times 2$ MIMO channel. Given the results inroduced above, we recall a construction method for fast-decodable space--time codes for this channel.  

\begin{theorem}\cite{vehkalahti:asymmetric}
	Let $\mc{C} = (K/\Q,\sigma,\gamma)$ be a division algebra of index $4$, where $K$ is a totally complex field containing a totally real field of index $2$. Assume that 
	\begin{itemize}
		\item[i)] $\left[K:\Q\right] = 4$,
		
		\item[ii)] $\gamma,\gamma^2 \not\in \nm{K/\Q}{K^\times}$,
		
		\item[iii)] $\gal(K/\Q) = \langle \sigma \rangle$ with $\sigma^2$ complex conjugation,
		
		\item[iv)] $\gamma < 0$. 
	\end{itemize}
	
	Let $\mc{O}_K = \Z w_1+\Z w_2+\Z w_3+\Z w_4$ be the ring of integers of $K$, and consider the left regular representation $\rho$ of $x \in \mc{C}$, which under the above assumptions can be written as 
	\begin{align*}
		\rho: x \mapsto \begin{bmatrix}
		x_1 & \gamma\sigma(x_4) & \gamma x_3^\ast & \gamma\sigma(x_2)^\ast \\
		x_2 & \sigma(x_1) & \gamma x_4^\ast & \gamma\sigma(x_3)^\ast \\
		x_3 & \sigma(x_2) & x_1^\ast & \gamma\sigma(x_4)^\ast \\
		x_4 & \sigma(x_3) & x_2^\ast & \sigma(x_1)^\ast
		\end{bmatrix}	
	\end{align*}
	Here, $x_i = g_{4i-3}w_1+g_{4i-2}w_2+g_{4i-1}w_3+g_{4i}w_4$ for $i = 1,\ldots,4$ with $g_j \in \Q$ for all $j$, and $\ast$ denotes complex conjugation. 
	
	For $\psi$ the explicit map given in Theorem~\ref{thm:embedding}, $\psi(\Gamma)$ is a lattice of dimension 16 in $\mat(4,\C)$ with the non-vanishing determinant property. For a signaling alphabet $S$, codes arising from this construction have a decoding complexity order of $10 \le k' \le 16$, that is, enjoy a reduction in decoding complexity of up to $37.5\%$.
\end{theorem}

\begin{example}
	The MIDO$_{A_4}$ code is a space--time code constructed in \cite{vehkalahti:asymmetric}. It is in fact a $(2,2,4)$ block orthogonal code, constructed from an algebra over the fifth cyclotomic field $\Q(\zeta_5)$. Consider the cyclic division algebra 
	\begin{align*}
		\mc{C} = \left(\Q(\zeta_5)/\Q,\sigma,-\frac{8}{9}\right),
	\end{align*} 
	where $\sigma:\zeta_5 \mapsto \zeta_5^3$. 
	
	Fix the $\Z$-basis $\left\{1-\zeta_5,\zeta_5-\zeta_5^2,\zeta_5^2-\zeta_5^3,\zeta_5^3-\zeta_5^4\right\}$ of $\mc{O}_K$. Consider a maximal order $\Gamma$ of $\mc{C}$, and $\psi$ the conjugation given in Theorem~\ref{thm:embedding}. Under this conjugation, codewords are of the form
	\begin{align*}
	X(x_1,\ldots,x_4) = \begin{bmatrix}
		x_1 & -r^2x_1^\ast & -r^3\sigma(x_4) & -r\sigma(x_3)^\ast \\
		r^2 x_2 & x_1^\ast & r\sigma(x_3) & -r^3\sigma(x_4)^\ast \\
		rx_3 & -r^3 x_3^\ast & \sigma(x_1) & -r^2\sigma(x_2)^\ast \\ 
		r^3 x_3 & r x_2^\ast & r^2\sigma(x_1) & \sigma(x_1)^\ast
		\end{bmatrix},		
	\end{align*}
	where $r = \left(\frac{8}{9}\right)^{1/4}$ and 
	\begin{align*}
		x_i &= g_{4i-3}(1-\zeta_5)+g_{4i-2}(\zeta_5-\zeta_5^2)+g_{4i-1}(\zeta_5^2-\zeta_5^3)+g_{4i}(\zeta_5^3-\zeta_5^4),\\
		\sigma(x_i) &= g_{4i-3}(1-\zeta_5^3)+g_{4i-2}(\zeta_5^3-\zeta_5)+g_{4i-1}(\zeta_5-\zeta_5^4)+g_{4i}(\zeta_5^4-\zeta_5^2).
	\end{align*}
	
	The decoding complexity order of this code is $k' = 12$, resulting in a reduction in decoding complexity of $25\%$. 
	
	By choosing the basis $\left\{1,\frac{\zeta_5+\zeta_5^{-1}}{2},\frac{\zeta_5-\zeta_5^{-1}}{2},\frac{\zeta_5^2-\zeta_5^{-2}}{4}\right\}$ of $\mc{O}_K$ instead, the decoding complexity can be further reduced. However, this is no longer an integral basis, and the price to pay is a smaller minimum determinant, yielding a slightly worse performance. 
\end{example}

\subsection{Distributed Space--Time Codes}
\label{subsec:distributed_stc}

The second setting we consider is a cooperative communications scenario. More concretely, we consider the communication of $(M+1)$ users with a single destination, where every user as well as the destination can be equipped with either a single antenna or multiple antennas. In this scenario, enabling cooperation and dividing the allocated transmission time allows for the $M$ inactive users to aid the active source in communicating with the destination by acting as intermediate \emph{relays}. For more details on the transmission model we refer to \cite{yang:af, barreal:fd_stc_relay}. While this is a more involved transmission scheme, from the destinations point of view it can be modeled as a virtual MIMO channel. Assume that the destination is equipped with $n_r$ receive antennas. Setting $T = n := 2Mn_t$, where $n_t$ is the number of transmit antennas available at each transmitter, we get the familiar channel equation $Y = HX + N$,
where $X \in \mat(n,\C)$ and $Y \in \mat(n_r\times n,\C)$ are the (overall) transmitted and received signals, and the structure of the channel matrix $H \in \mat(n_r\times n,\C)$ is determined by the different relay paths\footnote{As remarked in Section~\ref{subsec:ml}, the property of fast decodability is independent of the channel. Hence, we omit details on the structure of the effective channel.}.

Furthermore, it is discussed in \cite{yang:af} that for this channel model, block-diagonal space--time codes, that is, where each $X \in \mc{X}$ takes the form 
\begin{align*}
	X = \diag\left(X_m\right)_m = \begin{bmatrix} X_1 & & \\ & \ddots & \\ & & X_M \end{bmatrix}
\end{align*}
with $X_m \in \mat(2n_t, \C)$ are good choices for this channel if they additionally respect the usual design criteria such as non-vanishing determinants. To achieve this block structure, the following function is crucial.
\begin{definition}
\label{def:coop_construction}
Consider an $M$-relay channel as discussed above. Given a space--time code $\mc{X} \subset \mat(2n_t,\C)$ and a suitable function $\eta$ of order $M$ (\emph{i.e.}, $\eta^M(X) = X$), define the function
\begin{align*}
		\Psi_{\eta,M}: \mc{X} &\to \mat(2n_t M,\C) \\
		X &\mapsto \diag\left\{\eta^i(X)\right\}_{i=0}^{M-1} = \begin{bmatrix} X & & \\ & \ddots & \\ & & \eta^{M-1}(X) \end{bmatrix}.
\end{align*}
\end{definition}

\begin{figure}[!h]
\centering
	\includegraphics{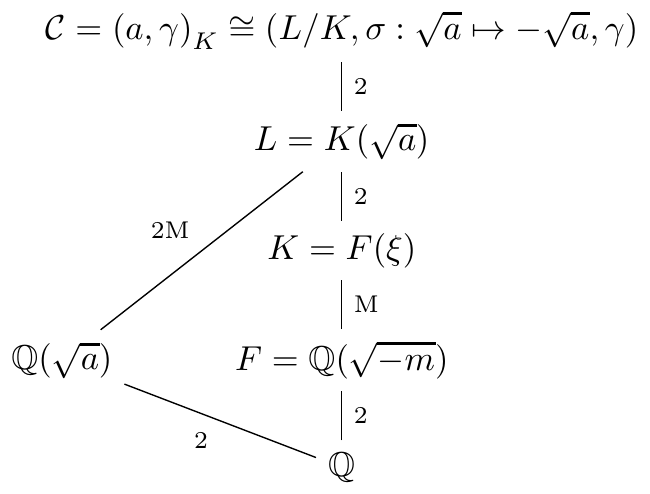}
\caption{Tower of extensions for the SIMO code construction.}
\label{fig:siso_tower}
\end{figure}

We begin with the case where $n_t = 1$ and $n_r \ge 2$. Consider the tower of extensions depicted in Figure~\ref{fig:siso_tower}, where $\xi$ is taken to be totally real, $m \in \Z_{\ge 1}$ and $a \in \Z\backslash\left\{0\right\}$ are square-free. 
Assume that $\mc{C}$ is division. Let $ \sigma$ be the generator of $ \gal(L/K)$, and fix a generator $ \eta $ of $ \gal(K/F)$. 

To have balanced energy and good decodability, it is necessary to slightly modify the matrix representation of the elements in $\mc{C}$, thus for $\Gamma \subset \mc{C}$ an order, instead of representing $x = x_0+\sqrt{\gamma} x_1 \in \Gamma$ by its left-regular representation $\rho(x)$, we define the following similar and commonly used function that maintains the original determinant,
\begin{equation}
\label{eqn:lambda_sim}
	\tilde{\rho}: x \mapsto \begin{bmatrix} x_0 & -\sqrt{-\gamma}\sigma(x_1) \\ \sqrt{-\gamma}x_1 & \sigma(x_0) \end{bmatrix}.
\end{equation}

\begin{theorem}\cite[Thm.~1]{barreal:fd_stc_relay}
\label{thm:single_antenna_relay_code}
Arising from the algebraic setup in Figure~\ref{fig:siso_tower} with $a < 0$, $\gamma < 0$, define the set

\begin{align*}
	\mc{X} = \left\{\Psi_{\eta,M}(X)\right\}_{X \in \tilde{\rho}(\Gamma)} = \left\{\left.\diag\left(\eta^i(X) \right)_{i=0}^{M-1} \right| X \in \tilde{\rho}(\Gamma) \right\}.
\end{align*}
	
The code $\mc{X}$ is of rank $8M$, rate $R = 4$ real symbols per channel use and has the non-vanishing determinant property. It is full-rate if $n_r = 2$. Moreover, $\mc{X}$ is conditionally $4$-group decodable, and its decoding complexity order can be reduced from $k = 8M$ to $k' = 5M$, resulting in a complexity reduction of $37.5\%$.
\end{theorem}

\begin{example}
For $M = 2$ relays and $\xi = \sqrt{5}$, consider the tower of extensions in Figure~\ref{fig:tower_siso}. 
\begin{figure}[h!]
	\centering
	\includegraphics[width=.6\textwidth]{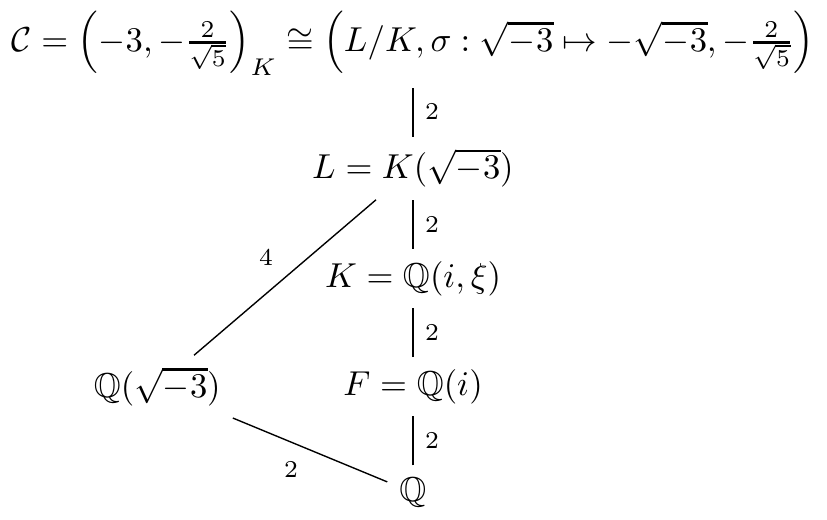}
	\caption{Tower of extensions for the SIMO example code.}
	\label{fig:tower_siso}
\end{figure}
The algebra $\mc{C}$ is division \cite[Exp.~1]{barreal:fd_stc_relay}. 
 
Let $x = x_0 + \sqrt{-\gamma}x_1$ with $x_0,x_1 \in \mathcal{O}_{L}$ and $X = \tilde{\rho}(x)$. 
For $\langle \eta \rangle = \Gamma(K/F)$, define the $2$-relay code
\begin{equation*}
	\mc{X} = \left\{ \Psi_{\eta,2}(X) \right\}_{X \in \tilde{\rho}(\mc{O}_{L})} = \left\{\left. \diag\left(\eta^i(X)\right)_{i=0}^{1} = \begin{bmatrix} X & \\ & \eta(X) \end{bmatrix} \right| X \in \tilde{\rho}(\mc{O}_{L}) \right\}. 
\end{equation*}

The resulting code is a fully diverse code of rank 16 with non-vanishing determinants, which is conditionally $4$-group decodable having decoding complexity oder $k' = {10}$ in contrast to $k = {16}$.  
\end{example}

\begin{figure}[!h]
\centering
	\includegraphics{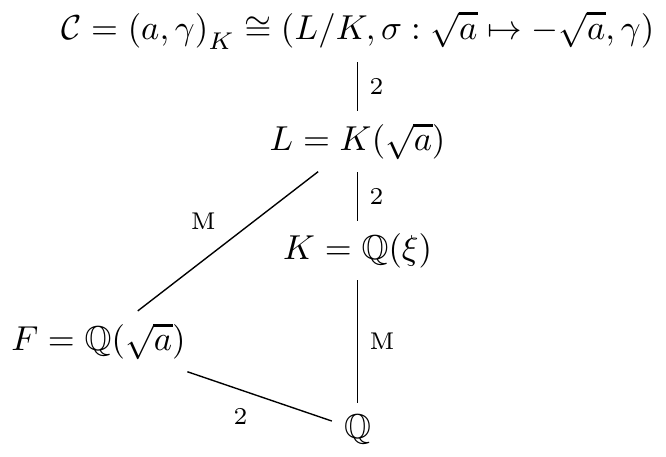}
\caption{Tower of extensions for the MIMO code construction.}
\label{fig:mimo_tower}
\end{figure}

We move on to the case where the transmitter and each relay is now equipped with $n_t \ge 1$ antennas. We require that the number of relays is expressible as $M  = (p-1)/2$, with $p \ge 5$ prime. Let henceforth $n_t = 2$. Assume further a single destination with $n_r \ge 1$ antennas, and consider the tower of extensions in Figure~\ref{fig:mimo_tower},
where $K = \Q(\xi) = \Q^+(\zeta_p) \subset \Q(\zeta_p)$ is the maximal real subfield of the $p^{\mathrm{th}}$ cyclotomic field, that is, $\xi = \zeta_p+\zeta_p^{-1}$, and $a \in \Z\backslash\left\{0\right\}$ is square-free. Let $\langle \sigma \rangle = \gal(L/K)$ and $\langle \eta \rangle = \gal(L/F)$. 

\begin{theorem}\cite[Thm.~2]{barreal:fd_stc_relay}
\label{thm:mult_antenna_relay_code}
	In the setup as in Figure~\ref{fig:mimo_tower}, choose $a \in \Z_{< 0}$ such that $\mf{p} = a\mc{O}_{K}$ is a prime ideal. Fix further $\gamma < 0$ and $\theta \in \mc{O}_{K}\cap \R^\times = \Z[\xi]\cap\R^\times$ such that 
	\begin{itemize}
		\item $\gamma$ and $\theta$ are both non-square $\bmod\ \mf{p}$,
		\item the quadratic form $\langle\gamma,-\theta\rangle_{L} := l_1\gamma-l_2\theta$ with $l_1,l_2 \in L$ is anisotropic, \emph{i.e.}, evaluates to zero if and only if $\gamma = \theta = 0$,
	\end{itemize}
and further let $\tau = \sigma$. Then, if $\Gamma \subset \mc{C}$ is an order, the distributed space--time code
\begin{align*}
	\mc{X} = \left\{\left.\Psi_{\eta,M}(\tilde{\alpha}_{\tau,\theta}(X,Y)) = \diag\left(\eta^i(\tilde{\alpha}_{\tau,\theta}(X,Y))\right)_{i=0}^{M-1} \right| X,Y \in \tilde{\rho}(\Gamma)\right\}
\end{align*}
is a full-diversity space--time code of rank $8M$, rate $R = 2$ real symbols per channel use (hence full-rate for $n_r = 1$)	, exhibits the non-vanishing determinant property and is $g$-group decodable, with $g \in \left\{2,4\right\}$. Its decoding complexity order is
\begin{align*}
	k' = \begin{cases} 4M&\mbox{if } a \equiv 1 \bmod\ 4, \\
	2M&\mbox{if } a \not\equiv 1 \bmod\ 4, 
	\end{cases}
\end{align*}
resulting in a reduction in complexity of $50\%$ and $75\%$, respectively.
\end{theorem}

\begin{example}
\label{exp:mimo_fd}
We construct a $4$-group decodable code for $M = 3$ relays, arising from the tower of extensions depicted in Figure~\ref{fig:tower_mimo}, where $\xi = \zeta_7+\zeta_7^{-1}$ and $\gamma = -\frac{2}{1+\xi}$.
\begin{figure}[h!]
\centering
	\includegraphics[width=0.6\textwidth]{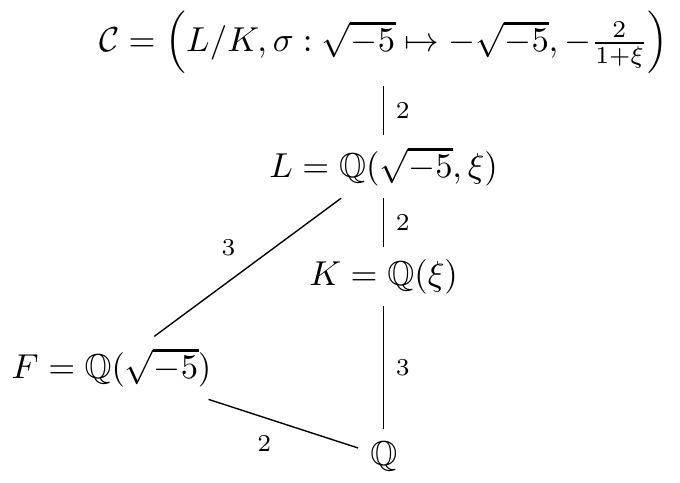}
	\caption{Tower of extensions for the MIMO example code.}
	\label{fig:tower_mimo}
\end{figure}  

In the following, let $\tau = \sigma$ and $\langle\eta:\xi\mapsto\xi^2-2\rangle = \Gamma\left(L/F\right)$. Choose further $\theta = 3(1-\xi) = \zeta\theta'$, with $\zeta = -1$ and $\theta' \in \R_{>0}$, and let $p_{\min}(x,\xi)$ be the minimal polynomial of $\xi$. With these choice of elements, the conditions from Theorem~\ref{thm:mult_antenna_relay_code} are satisfied. 

Let $x \in \Gamma \subset \mc{C}$, and set $\omega = \sqrt{-5}$. We define a space--time code $\mc{X}_{0}$ consisting of codewords of the form 
\begin{align*}	
	X = \tilde{\rho}(x) = \begin{bmatrix} x_{1} + x_{2}\omega & -\sqrt{-\gamma}(x_{3} + x_{4}\sigma(\omega)) \\ \sqrt{-\gamma}(x_{3} + x_{4}\omega) & x_{1} + x_{2}\sigma(\omega) \end{bmatrix},
\end{align*}
where $x_{i} \in \mc{O}_{K}$, $1 \le i \le 4$. Next, we iterate $\mc{X}_{0}$ with the help of $\tilde{\alpha}(\cdot,\cdot)$ to obtain the set
\begin{align*}
	\mc{X}_{0}^{\mathrm{it}} = \left\{\left.\tilde{\alpha}_{\tau,\theta}(X,Y) = \begin{bmatrix} X & \zeta\sqrt{\theta'}\tau(Y) \\ \sqrt{\theta'}Y & \tau(X) \end{bmatrix} \right| X,Y \in \tilde{\rho}(\Gamma) \right\}
\end{align*}
and finally adapt the iterated code to the 3-relay channel by applying the map $\eta$, resulting in distributed space--time code
\begin{align*}
	\mc{X} = \left\{\left.\Psi_{\eta,3}(\tilde{\alpha}_{\tau,\theta}(X,Y)) = \diag\left(\eta^j(\tilde{\alpha}_{\tau,\theta}(X,Y))\right)_{j = 0}^2 \right| X,Y \in \tilde{\rho}(\Gamma)\right\}
\end{align*}

The resulting relay code is fully diverse, exhibit the non-vanishing determinant property and are fast-decodable. More concretely, $\mc{X}$ is $4$-group decodable with decoding complexity order $k' = {6}$ in contrast to $k = {24}$, resulting in a complexity reduction of $75\%$. 
\end{example}
\newpage

\section*{Conclusions}
In this chapter we have given an overview on the topic of fast decodability of algebraic space--time codes. Traditionally, space--time codes have been developed in the context of point-to-point MIMO communications. However, with the development of new communication protocols in order to accommodate different types of applications and devices in modern wireless networks, so-called  distributed space--time codes have recently become a popular subject of research. Due to the nature of the underlying communication protocols, such codes often exhibit a too high decoding complexity for practical use. Following the ideas of fast-decodability of more traditional space--time codes, this chapter aimed at giving an overview on the subdivision of space--time codes into different families of so-called fast-decodable codes. Moreover, we were particularly interested in the specific reduction in decoding complexity offered by these codes. 

While crucial for practical implementation, only few explicit construction methods of fast-decodable space--time codes can be found in literature. In this chapter, we further recalled explicit constructions of asymmetric and distributed space--time codes with reduced decoding complexity, accompanied by example codes to illustrate the presented methods. 

With the upcoming fifth generation wireless systems in mind, the development of new constructions of suitable well-performing space--time codes which offer a reduction in decoding complexity is crucial for applications, and opens an interdisciplinary and rich research direction for future work.

\end{document}